\documentclass[12pt,a4paper]{article}
\pdfoutput=1
\usepackage{jheppub}
\usepackage{amsmath,amssymb,bbm,slashed,cancel}
\usepackage{graphicx}
\usepackage{slashed}
\usepackage{hyperref}
\usepackage{subcaption}
\usepackage{dsfont}
\usepackage{stackengine}
\usepackage[utf8]{inputenc}
\usepackage{multirow}
\usepackage{physics}
\usepackage{mathtools}

\usepackage{comment}

\usepackage{colortbl,colordvi,xcolor,color}

\usepackage{array,multirow}
\usepackage{booktabs}
\usepackage{float}

\allowdisplaybreaks

\newcommand{\eg}{{\em e.g.}}
\newcommand{\ie}{{\em i.e.}}
\newcommand\unit[1]{\,\mathrm{#1}}

\newcommand\GeV{\unit{GeV}}
\newcommand\TeV{\unit{TeV}}

\makeatletter
\gdef\@fpheader{}
\makeatother

\preprint{KEK--TH--2548}

\title{Higgs Probes of axion-like particles}
\author[a]{Masashi Aiko}
\author[a,b]{and Motoi Endo}

\affiliation[a]{KEK Theory Center, Tsukuba, Ibaraki 305--0801, Japan}
\affiliation[b]{Graduate Institute for Advanced Studies, SOKENDAI, Tsukuba, Ibaraki 305--0801, Japan}

\abstract{
We study axion-like particle contributions to the Higgs boson decays.
The particle is assumed to couple with the standard model electroweak gauge bosons.
Although direct productions of axion-like particles have often been discussed, we investigate indirect contributions to the Higgs boson decays into fermions, photons, $W$, and $Z$ bosons at the one-loop level. 
It is found that the corrections to the fermions are suppressed, whereas precise measurements of the di-photon channel of the Higgs boson decay can provide a significant probe of the model especially when the axion-like particle is heavy and its coupling to di-photon is suppressed.
}

\begin{document}
\maketitle
\flushbottom

\renewcommand{\thefootnote}{\#\arabic{footnote}}
\setcounter{footnote}{0}

\section{Introduction} \label{sec: intro}

The introduction of an axion-like particle (ALP) is one of the simplest extensions of the Standard Model (SM) and has been attracting wide interest. 
It is often motivated in models with spontaneous violations of global symmetries and is characterized by interactions with gauge bosons. 

We consider ALP primarily coupled with the SM ${\rm U(1)}_Y$ and ${\rm SU(2)}_L$ gauge bosons.
It then interacts with the photon ($\gamma$), $Z$ boson, and charged $W$ boson ($W^{\pm}$) after the electroweak (EW) symmetry breaking.
Such a particle affects experimental observables. 
For example, productions of ALPs at colliders have been studied extensively~\cite{Mimasu:2014nea, Jaeckel:2015jla, Jaeckel:2012yz, Bauer:2017ris, Bauer:2018uxu, Florez:2021zoo, Wang:2021uyb, dEnterria:2021ljz, Knapen:2016moh, CMS:2018erd, ATLAS:2020hii, Craig:2018kne, Bonilla:2022pxu}.
They are produced in heavy flavor decays if they are light enough~\cite{Izaguirre:2016dfi, Alonso-Alvarez:2018irt, Gavela:2019wzg, Guerrera:2021yss, Bauer:2021mvw, Guerrera:2022ykl}.
Moreover, they can affect electroweak precision observables through direct ALP production in the $Z$ boson decay or indirectly via vacuum polarizations of the EW gauge bosons~\cite{Bauer:2017ris, Aiko:2023trb}.

In this paper, we study ALP contributions to decay widths of the Higgs boson ($h$).
In literature, direct ALP productions from Higgs boson decays, \eg, $h \to Za$ and $h \to aa$, have often been discussed~\cite{Bauer:2017ris, Evans:2019xer, Davoudiasl:2021haa, Alves:2021puo, Cepeda:2021rql}.
Those channels are open when the ALP is lighter than the Higgs boson.
However, if it is heavier, the direct production channels are forbidden kinematically, and the ALP affects the Higgs boson decays via radiative corrections. 
In this paper, we will investigate ALP-loop contributions to the Higgs boson decays: $h \to f\bar f$ with $f$ denoting SM fermions, $h \to \gamma\gamma$, $h \to Z\gamma$, $h \to ZZ^*$, and $h \to WW^*$.
They will be compared with the direct ALP productions from Higgs boson decays. 
We will also take collider constraints into account.
It will be shown that the contribution to the fermion channels is suppressed, while $h \to \gamma\gamma$ provides a significant probe of the ALP compared with other channels such as $h \to Z\gamma$, $h \to ZZ^*$, $h \to WW^*$, and $h\to aa$.

This paper is organized as follows.
In Sec.~\ref{sec: ALP_model}, we introduce the ALP model. 
In Sec.~\ref{sec: decay_Higgs}, the ALP contributions to Higgs boson decays are explained. 
Numerical results are given in Sec.~\ref{sec: numerical results}, and Sec.~\ref{sec: conclusion} is devoted to the conclusion.
In Appendix \ref{app: 1PI}, the analytic expressions for the ALP contributions at the one-loop level are summarized.

\section{Model} \label{sec: ALP_model}

Let us introduce ALP $(a)$. 
It is assumed to couple with the ${\rm SU(2)}_L$ gauge boson $(W_\mu^a)$ and the ${\rm U(1)}_Y$ gauge boson $(B_\mu)$. 
We do not consider ALP couplings with other SM particles.
The Lagrangian is given by
\begin{align}
\mathcal{L}_{\rm ALP}
&=
\frac{1}{2}\partial_{\mu}a\partial^{\mu}a
-\frac{1}{2} m_{a}^{2} a^{2}
-c_{WW}\frac{a}{f_{a}}W_{\mu\nu}^{a}\widetilde{W}^{a\mu\nu}
-c_{BB}\frac{a}{f_{a}}B_{\mu\nu}\widetilde{B}^{\mu\nu},
\label{eq: Lagrangian}
\end{align}
where $m_a$ and $f_a$ are the mass and decay constant of the ALP, respectively.
In this paper, the coupling constants, $c_{WW}$ and $c_{BB}$, as well as $m_a$ and $f_a$ are regarded as free parameters but $m_a < f_a$ is satisfied. 
Field strengths of the gauge bosons and their dual are shown as
\begin{align}
W_{\mu\nu}^{a} &= \partial_{\mu}W_{\nu}^{a}-\partial_{\nu}W_{\mu}^{a}+g\epsilon^{abc}W_{\mu}^{b}W_{\nu}^{c}, \\
B_{\mu\nu} &= \partial_{\mu}B_{\nu}-\partial_{\nu}B_{\mu}, \\
\widetilde{X}_{\mu\nu} &= \frac{1}{2}\epsilon^{\mu\nu\rho\sigma}X_{\rho\sigma}\qc
(X = W^{a}, B)
\end{align}
where $g$ is the ${\rm SU(2)}_L$ gauge coupling constant.
The totally antisymmetric tensors are defined with $\epsilon^{012}=1$ and $\epsilon^{0123}=1$.
After the EW symmetry breaking, the above interactions are expressed as
\begin{align}
\mathcal{L}_{\mathrm{int}}
&=
-\frac{1}{4}g_{a\gamma\gamma}aF_{\mu\nu}\widetilde{F}^{\mu\nu}
-\frac{1}{2}g_{a\gamma Z}aZ_{\mu\nu}\widetilde{F}^{\mu\nu} \notag \\
&\quad
-\frac{1}{4}g_{aZZ}aZ_{\mu\nu}\widetilde{Z}^{\mu\nu}
-\frac{1}{2}g_{aWW}aW_{\mu\nu}^{+}\widetilde{W}^{-\mu\nu}
+ \cdots,
\label{eq: effective_coupling}
\end{align}
where quartic interaction terms are omitted, and the field strengths denote $X_{\mu\nu} = \partial_{\mu}X_{\nu}-\partial_{\nu}X_{\mu}$. 
Here and hereafter, $A_\mu$, $Z_\mu$, and $W_\mu$ represent the photon, $Z$, and charged $W$ bosons, respectively. 
Those coupling constants are related to $c_{WW}$ and $c_{BB}$ as
\begin{align}
g_{a\gamma\gamma} &= \frac{4}{f_{a}}\qty(s_{W}^{2}{c}_{WW}+c_{W}^{2}{c}_{BB}), 
\label{eq: gaAA} \\
g_{aZ\gamma} &= \frac{2}{f_{a}}\qty(c_{WW}-c_{BB})s_{2W}, 
\label{eq: gaZA} \\
g_{aZZ} &= \frac{4}{f_{a}}\qty(c_{W}^{2}c_{WW}+s_{W}^{2}c_{BB}), 
\label{eq: gaZZ} \\
g_{aWW} &= \frac{4}{f_{a}}c_{WW}.
\label{eq: gaWW}
\end{align}
Here and hereafter, $c_{W} = \cos{\theta_{W}}$, $s_{W} = \sin{\theta_{W}}$, $c_{2W} = \cos{2\theta_{W}}$, and $s_{2W} = \sin{2\theta_{W}}$ with the Weinberg angle $\theta_{W}$.
It is stressed that the original ALP interactions in our setup are governed by two coupling constants, ${c}_{BB}$ and ${c}_{WW}$, though there are four couplings, $g_{a\gamma\gamma}, g_{aZ\gamma}, g_{aZZ}$, and $g_{aWW}$, after the EW symmetry breaking. 
It is often useful to represent $g_{aZ\gamma}$ and $g_{aZZ}$ in terms of $g_{a\gamma\gamma}$ and $g_{aWW}$,
\begin{align}
g_{aZ\gamma} &= \frac{s_{W}(g_{aWW}-g_{a\gamma\gamma})}{c_{W}}, \label{eq: gaZgamma} \\
g_{aZZ} &= \frac{c_{2W}g_{aWW}+s_{W}^{2}g_{a\gamma\gamma}}{c_{W}^{2}}.
\end{align}

\section{Higgs boson decay} \label{sec: decay_Higgs}

In this section, we explain ALP contributions to the Higgs boson decays at the one-loop level. 
The decay channels in interest are $h \to f\bar f$, $h \to ZZ^{*}$, $h \to WW^{*}$, $h \to \gamma\gamma$, and $h \to Z\gamma$ as well as $h\to aa$.\footnote{
It is noted that there is no ALP contribution to $h\to gg$ at the one-loop level in our setup.
Also, $h\to Z a$ is not induced at this level because gauge-boson loop contributions vanish~\cite{Bauer:2017ris}.
}
Within the SM, the first three of them proceed at the tree level, and the following two are induced radiatively. 
Also, the last channel is absent.
The ALP affects all of them via radiative corrections.

For the radiative corrections, we adopt the on-shell renormalization to the EW sector by following Refs.~\cite{Bohm:1986rj, Hollik:1988ii}\footnote{Since there is no ALP mixing with the CP-odd Nambu-Goldstone boson at the one-loop level, we use the gauge fixing terms, which are the same as in the SM.
}.
The counterterms relevant to ALP corrections to the Higgs boson decays are introduced in Sec.~\ref{sec: renormalization}, and the decay width for each channel will be given in Sec.~\ref{subsec: hToff}, \ref{sec: hVV_vetex}, and \ref{sec: loopinduced_hVVprime_vetex}.

\subsection{Renormalization} \label{sec: renormalization}

The bare gauge fields (described with the subscript $B$) are related to the renormalized ones by means of the counterterms (denoted as $\delta X$) as
\begin{align}
W_{B \mu}^{\pm} &= \qty(1 + \frac{1}{2}\delta Z_{W})W_{\mu}^{\pm}, \\
\left(\begin{array}{c}
Z_{B \mu} \\
A_{B \mu}
\end{array}\right)
&=
\left(
\begin{array}{cc}
1+\cfrac{1}{2}\delta Z_{Z} & 
\cfrac{1}{2}\left(\delta Z_{Z \gamma} - \cfrac{\delta s_{W}^{2}}{c_{W}s_{W}}\right)\\
\cfrac{1}{2}\left(\delta Z_{Z \gamma}+\cfrac{\delta s_{W}^{2}}{c_{W}s_{W}}\right) & 
1+\cfrac{1}{2}\delta Z_{\gamma}
\end{array}
\right)
\left(\begin{array}{c}
Z_{\mu} \\
A_{\mu}
\end{array}\right).
\end{align}
The renormalized $W, Z$ masses and the measured value of the fine-structure constant are related to bare parameters as
\begin{align}
m_{W, B}^{2} = m_{W}^{2}+\delta m_{W}^{2}\qc
m_{Z, B}^{2} = m_{Z}^{2}+\delta m_{Z}^{2}\qc
\alpha_{{\rm em}, B} = \alpha_{\rm em}+\delta \alpha_{\rm em}.
\end{align}
The Weinberg angle and the Higgs vacuum expectation value (VEV) satisfy the following tree-level relations,
\begin{align}
s_{W, B}^{2} = 1-\cfrac{m_{W, B}^{2}}{m_{Z, B}^{2}}\qc
v_{B}^{2} = \cfrac{m_{W, B}^{2}s_{W, B}^{2}}{\pi \alpha_{\mathrm{em}, B}}.
\end{align}
Then, their counterterms are obtained as
\begin{align}
\frac{\delta s_{W}^{2}}{s_{W}^{2}} &= \frac{c_{W}^{2}}{s_{W}^{2}}\left(\frac{\delta m_{Z}^{2}}{m_{Z}^{2}}-\frac{\delta m_{W}^{2}}{m_{W}^{2}}\right), \label{eq: delta_sw}\\
\frac{\delta v}{v} &=
\frac{1}{2}\qty(
\frac{\delta m_{W}^{2}}{m_{W}^{2}}
+\frac{\delta s_{W}^{2}}{s_{W}^{2}}
-\frac{\delta \alpha_{\rm em}}{\alpha_{\rm em}}). \label{eq: delta_v}
\end{align}

Let us represent the renormalized inverse propagators of gauge bosons as
\begin{align}
\widehat{\Gamma}^{\mu\nu}_{VV'}
=
-ig^{\mu\nu}(p^{2}-m_{V}^{2})\delta_{VV'}
+i\widehat{\Pi}_{VV'}^{\mu\nu}(p^{2}),
\end{align}
with $V, V'= W, Z, \gamma$.
The self-energies $\widehat{\Pi}_{VV'}^{\mu\nu}(p^{2})$ are decomposed into longitudinal $(L)$ and transverse $(T)$ polarization components,
\begin{align}
\widehat{\Pi}_{VV'}^{\mu\nu}(p^{2}) =
\left(-g^{\mu\nu}+\frac{p^{\mu}p^{\nu}}{p^{2}}\right)\widehat{\Pi}_{VV'}^{T}(p^{2})
+\frac{p^{\mu}p^{\nu}}{p^{2}}\widehat{\Pi}_{VV'}^{L}(p^{2}).
\end{align}
Since the longitudinal component is irrelevant to the following discussion, we focus on the transverse one and omit the subscript, $T$ and $L$, for simplicity.\footnote{
Since the ALP is not coupled with the Higgs boson and fermions at the tree level, there are no ALP contributions to the Higgs and fermion self-energies.
}
In terms of the 1PI contributions and counterterms, they are given by\footnote{
Since there is no ALP contribution to tadpoles, we omit the tadpole terms in this paper.
}
\begin{align}
\widehat{\Pi}_{WW}(p^{2})
&=
\Pi_{WW}^{{\rm 1PI}}(p^{2})-\delta m_{W}^{2}+\delta Z_{W}(p^{2} -m _{W}^{2}), \\
\widehat{\Pi}_{ZZ}(p^{2})
&=
\Pi_{ZZ}^{{\rm 1PI}}(p^{2})-\delta m_{Z}^{2}+\delta Z_{Z}(p^{2}-m_{Z}^{2}), \\
\widehat{\Pi}_{\gamma\gamma}(p^{2})
&=
\Pi_{\gamma\gamma}^{{\rm 1PI}}(p^{2})+\delta Z_{\gamma}p^{2}, \\
\widehat{\Pi}_{Z\gamma}(p^{2})
&=
\Pi_{Z\gamma}^{{\rm 1PI}}(p^{2})+\delta Z_{Z\gamma}\left(p^{2}-\frac{1}{2}m_{Z}^{2}\right)
+\frac{m_{Z}^{2}}{2}\frac{\delta s_{W}^{2}}{c_{W}s_{W}}.
\end{align}
In the renormalization scheme explored by Refs.~\cite{Bohm:1986rj, Hollik:1988ii}, the residue of the two-point function for the weak gauge bosons is not unity.
Hence, we need to add a derivative of the self-energy, $\widehat{\Pi}_{VV}(m_{V}^{2})^{\prime}$, to transition amplitudes with external $W^{\pm}$ and $Z$ bosons, 
\begin{align*}
\widehat{\Pi}_{VV}(m_{V}^{2})^{\prime}
&=
\Pi_{VV}^{\rm 1PI}(m_{V}^{2})^{\prime}+\delta Z_{V}\qc (V= W^{\pm}, Z).
\end{align*}

Let us focus on the ALP contributions. 
They arise via $\Pi_{VV'}^{\rm 1PI}$, and the results are obtained in Appendix~\ref{app: two-point}.
In the following, the counterterms denote those from ALP, $\delta X = \delta X_{\rm ALP}$.
In addition, we use the relation, $\Pi_{VV'}^{\rm 1PI}(0)_{\rm ALP}=0$ (see Appendix~\ref{app: two-point}) to simplify the expressions.
Then, we obtain
\begin{align}
\delta m_{W}^{2} &= \Re \qty[\Pi^{{\rm 1PI}}_{WW}(m_{W}^{2})_{\rm ALP}], \\
\delta m_{Z}^{2} &= \Re \qty[\Pi^{{\rm 1PI}}_{ZZ}(m_{Z}^{2})_{\rm ALP}], \\
\frac{\delta \alpha_{\rm em}}{\alpha_{\rm em}}
&=
\Pi_{\gamma\gamma}^{{\rm 1PI}}(0)^{\prime}_{\rm ALP},
\end{align}
yielding $\delta s_{W}^{2}$ and $\delta v$ (see Eqs.~\eqref{eq: delta_sw} and \eqref{eq: delta_v}).
The wave function renormalization constants are shown as
\begin{align}
\delta Z_{\gamma} &= -\frac{\delta \alpha_{\rm em}}{\alpha_{\rm em}}\qc
\delta Z_{Z \gamma} = \frac{\delta s_{W}^{2}}{c_{W}s_{W}}, \label{eq: del_ZAV}\\
\delta Z_{V}
&=
2\frac{\delta v}{v}-\frac{\delta m_{V}^{2}}{m_{V}^{2}}\qc (V= W^{\pm}, Z).
\end{align}

In our calculation, the Higgs VEV $v$ is represented by the measured value of the Fermi coupling constant $G_{F}$,
\begin{align}
v^{2} = \frac{1}{\sqrt{2}G_{F}(1-\Delta r)},
\end{align}
where $\Delta r$ is obtained from the muon decay rate evaluated at the one-loop level~\cite{Sirlin:1980nh}.
In our setup, it satisfies the following relation,
\begin{align}
\Delta r = -2\frac{\delta v}{v}.
\label{eq: delta_r}
\end{align}

\subsection{$h\to f\bar{f}$} \label{subsec: hToff}

\begin{figure}[t]
 \centering
 \includegraphics[scale=0.45]{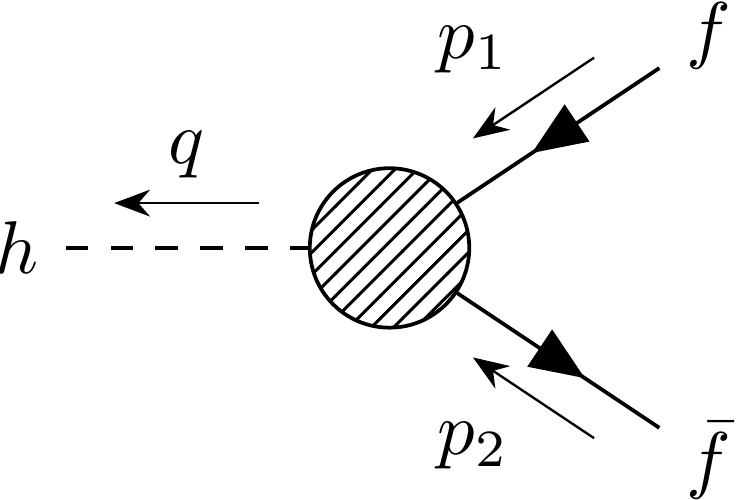} \hspace{7mm}
  \includegraphics[scale=0.45]{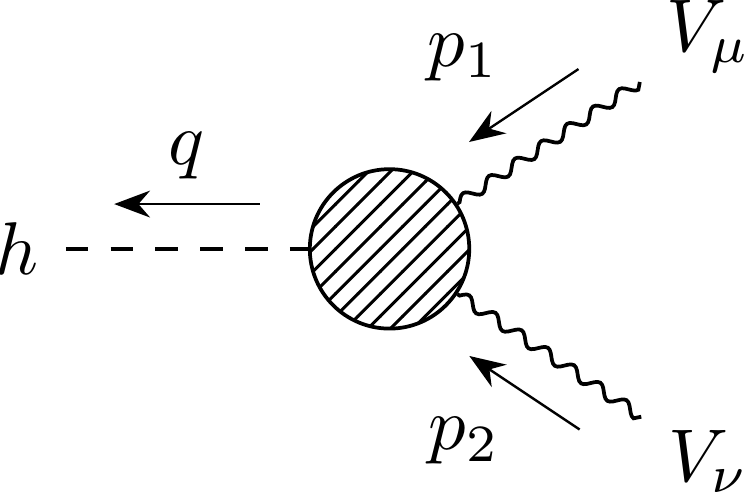}
 \caption{Momentum assignments to the Higgs boson decays.}
 \label{fig: mom_hff_hVV}
\end{figure}

The decay proceeds at the tree level in the SM, and the ALP affects it via radiative corrections. 
In general, the renormalized $hf\bar{f}$ vertex is decomposed as
\begin{align}
\widehat{\Gamma}_{hf\bar{f}}(p_{1}, p_{2}, q) &=
\widehat{\Gamma}_{hf\bar{f}}^{S}+\gamma_{5}\widehat{\Gamma}_{hf\bar{f}}^{P}
+\slashed{p}_{1}\widehat{\Gamma}_{hf\bar{f}}^{V_{1}}
+\slashed{p}_{2}\widehat{\Gamma}_{hf\bar{f}}^{V_{2}} \notag\\
&\quad
+\slashed{p}_{1}\gamma_{5}\widehat{\Gamma}_{hf\bar{f}}^{A_{1}}
+\slashed{p}_{2}\gamma_{5}\widehat{\Gamma}_{hf\bar{f}}^{A_{2}}
+\slashed{p}_{1}\slashed{p}_{2}\widehat{\Gamma}_{hf\bar{f}}^{T}
+\slashed{p}_{1}\slashed{p}_{2}\gamma_{5}\widehat{\Gamma}_{hf\bar{f}}^{PT}, \label{eq: hff-form}
\end{align}
where $p_{1}\ (p_{2})$ denotes the incoming four-momentum of the fermion (anti-fermion), and $q$ is the outgoing four-momentum of the Higgs boson (see the left panel of Fig.~\ref{fig: mom_hff_hVV}).
The form factors are composed of the tree and one-loop contributions as
\begin{align}
\widehat{\Gamma}_{hf\bar{f}}^{X} =
{\Gamma}_{hf\bar{f}}^{X,\mathrm{tree}}+{\Gamma}_{hf\bar{f}}^{X,\mathrm{loop}}\qc
(X = S, P, V_{1}, V_{2}, A_{1}, A_{2}, T, PT).
\end{align}
The former is given by
\begin{align}
\Gamma_{hf\bar{f}}^{S,{\rm tree}}=-\frac{m_{f}}{v}\qc
\Gamma_{hf\bar{f}}^{X,{\rm tree}}=0\quad (X\neq S).
\end{align}
The latter is composed of the 1PI and counterterm contributions,
\begin{align}
\Gamma^{X,{\rm loop}}_{hf\bar{f}}=\Gamma^{X,{\rm 1PI}}_{hf\bar{f}}+\delta \Gamma^{X}_{hf\bar{f}}.
\end{align}
In our setup, there is no ALP contribution to the 1PI terms at the one-loop level, and the counterterms do not vanish only for $\delta\Gamma^{S}_{hf\bar{f}}$ in the presence of the Higgs VEV counterterm.
Thus, the ALP contributions are obtained as
\begin{align}
\Gamma_{hf\bar{f},\, {\rm ALP}}^{S, {\rm loop}}
= 
\delta \Gamma_{hf\bar{f}}^{S} = -\Gamma_{hf\bar{f}}^{S,{\rm tree}}\frac{\delta v}{v}\qc
\Gamma_{hf\bar{f},\, {\rm ALP}}^{X, {\rm loop}} = 0\quad (X\neq S).
\label{eq: hff_loop}
\end{align}
As a result, the decay rate for $h\to f\bar{f}$ $(f \neq t)$ is expressed as
\begin{align}
\Gamma(h\to f\bar{f}) =
\Gamma_{\mathrm{LO}}(h\to f\bar{f}) \qty[1+\Delta_{\mathrm{SM}}^{f}+\Delta_{\mathrm{ALP}}^{f}]. 
\end{align} 
At the leading order, the rate is obtained at the tree level as
\begin{align}
\Gamma_{\mathrm{LO}}(h\to f\bar{f}) = N_{c}^{f}\frac{G_{F}m_{h}m_{f}^{2}}{4\sqrt{2}\pi}\lambda^{3/2}\qty(\frac{m_{f}^{2}}{m_{h}^{2}}, \frac{m_{f}^{2}}{m_{h}^{2}}),
\end{align}
with $N_{c}^{f}=3\,(1)$ for $f$ being quarks (leptons). 
The kinetic factor is defined as
\begin{align}
\lambda\qty(x, y) = (1-x-y)^2-4xy.
\end{align}
The SM correction $\Delta_{\rm{SM}}^f$ is found in Refs.~\cite{Kniehl:1991ze, Dabelstein:1991ky}, and the ALP term $\Delta_{\rm{ALP}}^f$ at the one-loop level is obtained from Eqs.~\eqref{eq: delta_r} and \eqref{eq: hff_loop} as
\begin{align}
\Delta_{\text{ALP}}^{f}
&=
\frac{2}{\abs{\Gamma_{hf\bar{f}}^{S,{\rm tree}}}^{2}}\Re
\qty[\Gamma_{hf\bar{f}}^{S,{\rm tree}}\qty(
\Gamma^{S, {\rm loop}}_{hf\bar{f},\, {\rm ALP}}
)^{*}]
-\Delta r 
=0.
\label{eq:delew}
\end{align}
Therefore, $\Delta_{\rm{ALP}}^f$ is canceled out, and there is no correction from ALP to $h\to f\bar{f}$ at this level.

\subsection{$h \to VV^{*}~(V=Z,W^{\pm})$} \label{sec: hVV_vetex}

The decay proceeds at the tree level in the SM, and the ALP affects it via radiative corrections. 
In general, the renormalized $hVV$ vertex is decomposed as
\begin{align}
\widehat{\Gamma}_{hVV}^{\mu\nu}(p_{1}, p_{2}, q) &= \qty[
g^{\mu\nu}\widehat{\Gamma}_{hVV}^1
+\frac{p_1^\nu p_2^\mu}{m_V^2}\widehat{\Gamma}_{hVV}^2
+i\epsilon^{\mu\nu\rho\sigma}\frac{p_{1\rho} p_{2\sigma}}{m_V^2}\widehat{\Gamma}_{hVV}^3
](p_{1}^{2}, p_{2}^{2}, q^{2}),
\end{align}
where $p_{1}^{\mu}$ and $p_{2}^{\mu}$ are the incoming four-momenta of the weak bosons, and $q^{\mu}$ is the outgoing four-momentum of the Higgs boson (see the right panel of Fig.~\ref{fig: mom_hff_hVV}).
The form factors are composed of the tree and one-loop contributions as
\begin{align}
\widehat{\Gamma}^X_{hVV}={\Gamma}^{i,{\rm tree}}_{hVV}+{\Gamma}^{i,{\rm loop}}_{hVV}\qc (i=1,2,3).
\end{align}
The former is given by
\begin{align}
\Gamma^{1, \mathrm{tree}}_{hVV}= \frac{2m_{V}^{2}}{v}\qc
\Gamma^{2, \mathrm{tree}}_{hVV}=\Gamma^{3, \mathrm{tree}}_{hVV}=0.
\end{align}
The latter is decomposed into the 1PI and counterterm contributions as\footnote{
The ALP contribution to the tadpole is absent and omitted in the expression.
}
\begin{align}
{\Gamma}^{i, \mathrm{loop}}_{hVV} = \Gamma^{i, \mathrm{1PI}}_{hVV}+\delta \Gamma^{i}_{hVV}.
\end{align}
The ALP contributes to $\Gamma^{1, \mathrm{1PI}}_{hVV}$ and $\Gamma^{2, \mathrm{1PI}}_{hVV}$ at the one-loop level.
The results are given in Appendix~\ref{app: three-point}.
The ALP contributions to the counterterms become non-zero only for $\delta \Gamma^{1}_{hVV}$ in the presence of the Higgs VEV counterterm as
\begin{align}
\delta \Gamma^{1}_{hVV} = \Gamma^{1, \mathrm{tree}}_{hVV}\frac{\delta v}{v}\qc
\delta \Gamma^{2}_{hVV} = \delta \Gamma^{3}_{hVV} = 0.
\end{align}

\subsubsection{$h \to ZZ^{*}$ \label{sec:hzz}}

Since the Higgs boson mass is smaller than $2m_{Z}$, one of the final-state $Z$ bosons is off-shell and decays into a pair of SM fermions. 
The decay rate for $h\to ZZ^\ast \to Z f\bar{f}$ is written as  
\begin{align}
\Gamma(h\to Z f\bar{f}) = \Gamma_{\mathrm{LO}}(h\to Z f\bar{f})\qty[1+\Delta_{\rm SM}^{Z}+\Delta_{\rm ALP}^{Z}].
\end{align}
In the following, we neglect masses of the fermions.
At the leading order, the rate is obtained at the tree level as
\begin{align}
\Gamma_{\mathrm{LO}}(h\to Z f\bar{f})
&=
N_{c}^{f}\int_0^{s_{\rm max}^{Z}}|\overline{{\cal M}_0^Z}|^2\,\dd s,
\end{align}
with $s_{\rm max}^{Z} = (m_h-m_Z)^2$.
The squared amplitude $|\overline{{\cal M}_{0}^Z}|^2$ is shown as 
\begin{align}
|\overline{{\cal M}_0^Z}|^2 = \frac{g_{Z}^{2}}{256\pi^3m_h^3} \abs{\Gamma_{hZZ}^{1,\text{tree}}}^2 \frac{v_f^2 + a_f^2}{(x_s - x_Z)^2}\frac{\lambda(x_Z,x_s)+12 x_Z x_s}{3x_Z}\lambda^{1/2}(x_Z,x_s), 
\label{eq:mZ0sq}
\end{align}
where $x_{Z} = m_{Z}^{2}/m_{h}^{2}$, $x_{s} = s/m_{h}^{2}$, $s = (p_{f} + p_{\bar{f}})^2$, $v_f = I_{3}^{f}/2-Q_{f}s_{W}^{2}$, and $a_f = I_{3}^{f}/2$ with the weak isospin $I_3^f$ and the electric charge $Q_f$.
The radiative corrections $\Delta_{\rm SM/ALP}^{Z}$ can be written in terms of the renormalized quantities~\cite{Kniehl:1993ay, Kanemura:2019kjg}.
The ALP correction $\Delta_{\rm ALP}^{Z}$ is obtained as\footnote{In general, there are also corrections from renormalized $Zf\bar{f}$ and $hf\bar{f}$ vertices and 1PI box diagrams (see Refs.~\cite{Kniehl:1993ay, Kanemura:2019kjg}). 
However, they do not receive ALP contributions in our setup and are omitted in the expression.}
\begin{align}
\Delta_{\rm ALP}^{Z} &=
\frac{2}{\Gamma_{\mathrm{LO}}}\int_0^{s_{\rm max}^{Z}}\!\dd s\, |\overline{{\cal M}_0^Z}|^2 \Re\Bigg\{
\left[ \frac{\Gamma_{hZZ}^{1,{\rm loop}}}{\Gamma_{hZZ}^{1,{\rm tree}}} + \frac{\bar{\lambda}(x_Z,x_s)}{x_Z^{}} \frac{\Gamma_{hZZ}^{2,{\rm loop}} }{\Gamma_{hZZ}^{1,{\rm tree}}}\right](m_Z^2,s,m_h^2) \notag \\
&\quad
+\frac{v_fQ_fc_Ws_W}{v_f^2+a_f^2}\frac{s - m_Z^2}{s} \left[\frac{\Gamma_{hZ\gamma}^{1, {\rm 1PI}}}{\Gamma_{hZZ}^{1,{\rm tree}}} 
+\bar{\lambda}(x_Z, x_s)\frac{m_{h}^{2}\Gamma_{hZ\gamma}^{2, {\rm 1PI}}}{\Gamma_{hZZ}^{1,{\rm tree}}} \right](m_Z^2,s,m_h^2) \notag\\
&\quad
-\frac{\widehat{\Pi}_{ZZ}(s)}{s-m_Z^2} - \frac{v_fQ_fs_W^{}c_W^{}}{v_f^2 + a_f^2}\frac{\widehat{\Pi}_{Z\gamma}(s)}{s} 
-\Delta r - \frac{1}{2}\widehat{\Pi}_{ZZ}(m_{Z}^{2})^{\prime}
\Bigg\}.
\label{eq:del_ew_z}
\end{align}
The kinematic factor $\bar{\lambda}(x,y)$ is defined by
\begin{align}
\bar{\lambda}(x,y) = \frac{1- x - y}{2}\frac{\lambda(x,y)}{\lambda(x,y) +12 x y}. 
\end{align}

\subsubsection{$h \to WW^{*}$}

Similarly to $h \to ZZ^{*}$, one of the final-state $W$ bosons is off-shell and decays into a pair of SM fermions. 
The decay rate for $h \to WW^{*} \to Wf\bar{f^{\prime}}$ is given by
\begin{align}
\Gamma(h\to W f\bar{f^{\prime}}) &= \Gamma_{\mathrm{LO}}(h\to Wf\bar{f^{\prime}})\qty[1+\Delta_{\rm SM}^{W} +\Delta_{\rm ALP}^{W}].
\end{align}
In the following, we neglect masses of the fermions.
At the leading order, the rate is obtained at the tree level as
\begin{align}
\Gamma_{\mathrm{LO}}(h\to Wf\bar{f^{\prime}})
&=
N_{c}^{f}\int_0^{s_{\rm max}^{W}}\!|\overline{{\cal M}_{0}^{W}}|^2\,\dd s,
\end{align}
with $s_{\rm max}^{W}=(m_{h}-m_{W})^{2}$.
The squared tree-level amplitude $|\overline{{\cal M}_{0}^{W}}|^2$ is obtained as
\begin{align}
|\overline{{\cal M}_{0}^{W}}|^2 =  \frac{g^{2}}{512\pi^3m_h^3}
\abs{\Gamma_{hWW}^{1,\text{tree}}}^2
\frac{\lambda(x_{W}, x_{s})+12 x_{W} x_{s}}{3x_{W}(x_{s} - x_{W})^2}\lambda^{1/2}(x_{W}, x_{s}), 
\label{eq:mW0sq}
\end{align}
with $x_{W} = m_{W}^{2}/m_{h}^{2}$ and $s = (p_{f} + p_{\bar{f^{\prime}}})^2$.
The radiative corrections $\Delta_{\rm SM/ALP}^{W}$ can be written in terms of the renormalized quantities~\cite{Kanemura:2019kjg}.
The ALP correction $\Delta_{\rm{ALP}}^{W}$ is obtained as\footnote{Similarly to $h\to ZZ^{*}$, there are generally corrections from renormalized $Wf\bar{f'}$ and $hf\bar{f}$ vertices and 1PI box diagrams. However, ALP contributions vanish in our setup, and only non-zero terms are shown.}
\begin{align}
\Delta_{\rm ALP}^{W} & =
\frac{2}{\Gamma_{\mathrm{LO}}}\int_0^{s_{\rm max}^{W}}\!\dd s\, |\overline{{\cal M}_0^W}|^2\Re\Bigg\{
\left[ \frac{\Gamma_{hWW}^{1,{\rm loop}}}{\Gamma_{hWW}^{1,{\rm tree}}} + \frac{\bar{\lambda}(x_W,x_s)}{x_W^{}} 
\frac{\Gamma_{hWW}^{2,{\rm loop}} }{\Gamma_{hWW}^{1,{\rm tree}}}\right](m_W^2,s,m_h^2) \notag \\ 
&\qquad\qquad\qquad\qquad\qquad\qquad\quad
-\frac{\widehat{\Pi}_{WW}(s)}{s-m_W^2}-\Delta r -\frac{1}{2}\widehat{\Pi}_{WW}(m_W^2)^{\prime}
\Bigg\}.
\label{eq:del_ew_w}
\end{align}
Note that since there is no photon-mediated diagram, we have no infrared (IR) divergences in ALP contributions. 
Thus, the treatment of IR regularization in the numerical analysis is the same as that in the SM~\cite{Kanemura:2019kjg}.

\begin{figure}[t]
 \centering
 \includegraphics[scale=0.45]{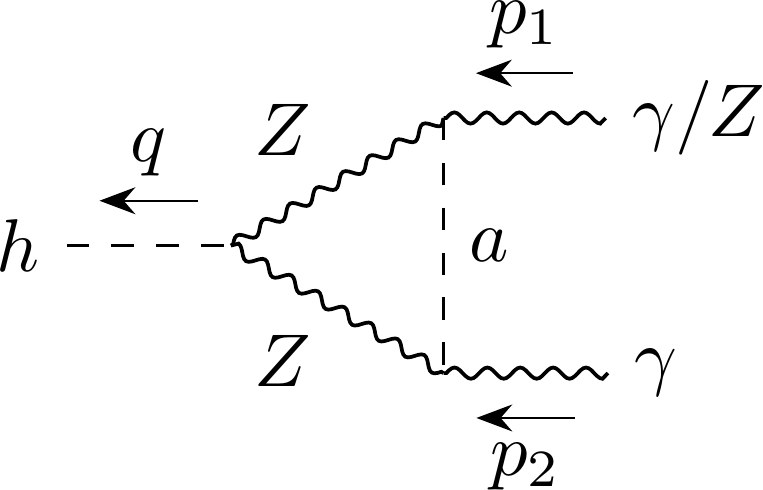} \hspace{5mm}
 \includegraphics[scale=0.45]{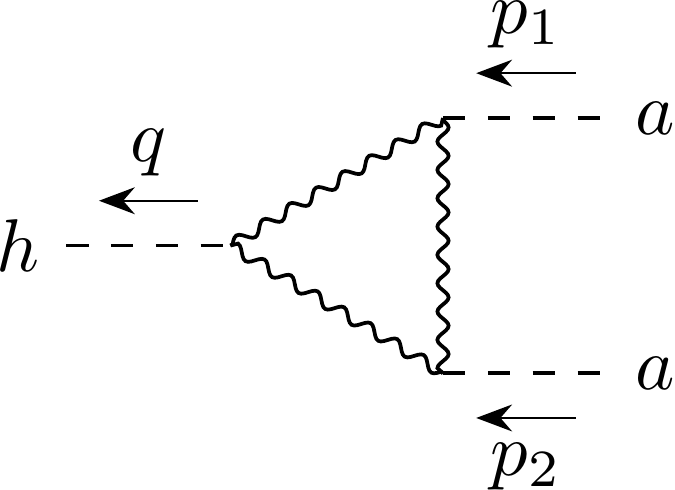}
 \caption{Feynman diagram of ALP contributions to $h \to \gamma\gamma, Z\gamma$ and $h \to aa$.
 In the right panel, the gauge bosons denote $W/Z/\gamma$.}
 \label{fig: hToGamGam}
\end{figure}

\subsection{$h \to V\gamma$\ $(V=\gamma, Z)$} \label{sec: loopinduced_hVVprime_vetex}

In contrast to the previous channels, $h \to V\gamma$ proceeds via radiative corrections in the SM as well as the ALP model. 
The renormalized $h\mathcal{V}\mathcal{V'}$ vertex is expressed as
\begin{align}
\widehat{\Gamma}_{h\mathcal{V}\mathcal{V'}}^{\mu\nu}(p_{1}, p_{2}, q) &= \qty[
g^{\mu\nu}\widehat{\Gamma}_{h\mathcal{V}\mathcal{V'}}^{1}
+p_{1}^{\nu}p_{2}^{\mu}\widehat{\Gamma}_{h\mathcal{V}\mathcal{V'}}^{2}
+i\epsilon^{\mu\nu\rho\sigma}p_{1\rho}p_{2\sigma}\widehat{\Gamma}_{\mathcal{V}\mathcal{V'}}^{3}
](p_{1}^{2}, p_{2}^{2}, q^{2}),
\end{align}
with $\mathcal{V}=\gamma, Z$ and $\mathcal{V'}=\gamma$. 
Here, $p_{1}^{\mu}$ and $p_{2}^{\mu}$ are the incoming four-momenta of the gauge bosons, and $q^{\mu}$ is the outgoing four-momentum of the Higgs boson (see the left panel of Fig.~\ref{fig: hToGamGam}).
The form factors are composed only of the one-loop contributions, which are generally decomposed into the 1PI and counterterm components as
\begin{align}
\widehat{\Gamma}^X_{h\mathcal{V}\mathcal{V'}}={\Gamma}^{i,{\rm loop}}_{h\mathcal{V}\mathcal{V'}}
=\Gamma^{i, \mathrm{1PI}}_{h\mathcal{V}\mathcal{V'}}+\delta \Gamma^{i}_{h\mathcal{V}\mathcal{V'}}\qc
(i=1,2,3).
\end{align}
Since the $Z\text{--}\gamma$ mixing cannot induce $h \to \gamma\gamma$ at the one-loop level, the counterterms for the $h\gamma\gamma$ vertex are zero.
However, those for $hZ\gamma$ generally exist as
\begin{align}
\delta \Gamma^{1}_{hZ\gamma} = \frac{\Gamma_{hZZ}^{1, {\rm tree}}}{2}\qty(\delta Z_{Z\gamma}-\frac{\delta s_{W}^{2}}{s_{W}c_{W}})\qc
\delta \Gamma^{2}_{hZ\gamma} = \delta \Gamma^{3}_{hZ\gamma} = 0.
\end{align}
In the present setup, it is found from Eq.~\eqref{eq: del_ZAV} that the ALP contribution to $\delta \Gamma^{1}_{hZ\gamma}$ is canceled out.
Therefore, the ALP contributions are obtained by the 1PI terms as
\begin{align}
\widehat{\Gamma}^X_{h\mathcal{V}\mathcal{V'}}=\Gamma^{i, \mathrm{1PI}}_{h\mathcal{V}\mathcal{V'}}\qc (i=1,2,3).
\end{align}
The results are obtained in Appendix~\ref{app: three-point}.

\subsubsection{$h \to \gamma\gamma$}
The decay rate for $h\to \gamma\gamma$ is given by 
\begin{align}
&\Gamma(h \to\gamma \gamma)
=
\frac{1}{16\pi m_{h}}\abs{\widehat{\Gamma}_{h\gamma\gamma}^{1}(0, 0, m_{h}^{2})}^{2} \notag \\
&=
\frac{\sqrt{2}G_{F}\alpha_{\mathrm{em}}^{2}m_{h}^{3}}{256\pi^{3}}
\abs{I_{W}(\tau_{W})
+\sum_{f}N_{c}^{f}Q_{f}^{2}I_{F}(\tau_{f})(1+\Delta_{\rm QCD}^{f})
+I_{\rm ALP}(\tau_{a})}^{2}, 
\end{align}
with $\tau_{i} = m_{h}^{2}/(4m_{i}^{2})$ for $i=W, f, a$.
The SM loop functions are given by~\cite{Ellis:1975ap, Shifman:1979eb, Djouadi:2005gi}
\begin{align}
I_{W}(\tau)
&=
\qty[2\tau^{2}+3\tau+3(2\tau-1)f(\tau)]\tau^{-2},
\label{eq: Wloop_hgamgam}\\
I_{F}(\tau)
&=
-2\qty[\tau+(\tau-1)f(\tau)]\tau^{-2}.
\label{eq: Floop_hgamgam}
\end{align}
with $f(\tau)$ defined as
\begin{align}
f(\tau) =
\begin{dcases}
\arcsin^{2}(\sqrt{\tau}) &(\tau \leq 1), \\
-\frac{1}{4}\qty[\ln{\frac{1+\sqrt{1-\tau^{-1}}}{1-\sqrt{1-\tau^{-1}}}}-i\pi]^{2} \quad &(\tau>1).
\end{dcases}
\label{eq: f_tau}
\end{align}
The ALP loop function is given by
\begin{align}
I_{\rm ALP}(\tau)
&=
\frac{4\pi v}{\alpha_{\rm em}m_{h}^{2}}\Gamma_{h\gamma\gamma}^{1, {\rm 1PI}}(0,0,m_{h}^{2})_{\rm ALP}.
\label{eq: ALPloop_hgamgam}
\end{align}
The term $\Delta_{\rm QCD}^{f}$ denotes QCD corrections to the quark-loop contributions.
We include it up to next-to-next-to-leading order, and whose expressions in the large top-mass limit ($\tau_{t}\ll 1)$ are given in Refs.~\cite{Dawson:1993qf, Spira:1995rr}.

\subsubsection{$h \to Z\gamma$}
The decay rate for $h\to Z\gamma$ is given by 
\begin{align}
&\Gamma_{\mathrm{LO}}(h \to Z\gamma)
=
\frac{1}{8\pi m_{h}}\qty(1-\frac{m_{Z}^{2}}{m_{h}^{2}})\abs{\widehat{\Gamma}_{hZ\gamma}^{1}(m_{Z}^2, 0, m_{h}^{2})}^{2} \notag \\
&=
\frac{\sqrt{2}G_{F}\alpha_{\mathrm{em}}^{2}m_{h}^{3}}{128\pi^{3}}
\qty(1-\frac{m_{Z}^{2}}{m_{h}^{2}})^{3} \notag \\
&\quad\times
\abs{J_{W}(\tau_{W}, \lambda_{W})
+\sum_{f}N_{c}^{f}Q_{f}v_{f}J_{F}(\tau_{f}, \lambda_{f})(1+\Delta_{\rm QCD}^{f})
+J_{\rm ALP}(\tau_{a}, \lambda_{a})
}^{2},
\end{align}
with $\tau_{i} = 4m_{i}^{2}/m_{h}^{2}$ and $\lambda_{i}=4m_{i}^{2}/m_{Z}^{2}$.
The SM loop functions are given by~\cite{Djouadi:2005gi, Cahn:1978nz, Bergstrom:1985hp}
\begin{align}
J_{W}(\tau, \lambda)
&=
\frac{c_{W}}{s_{W}}\bigg\{
\qty[\qty(1+\frac{2}{\tau})\frac{s_{W}^{2}}{c_{W}^{2}}-\qty(5+\frac{2}{\tau})]I_{1}(\tau, \lambda)
\notag \\ &\quad
+4\qty(3-\frac{s_{W}^{2}}{c_{W}^{2}})I_{2}(\tau, \lambda)
\bigg\},
\label{eq: Wloop_hZgam} \\
J_{F}(\tau, \lambda)
&=
\frac{4}{s_{W}c_{W}}\qty[I_{1}(\tau, \lambda)-I_{2}(\tau, \lambda)],
\label{eq: Floop_hZgam}
\end{align}
with $I_{1}(\tau, \lambda)$ and $I_{2}(\tau, \lambda)$ defined as
\begin{align}
I_{1}(\tau, \lambda) &= \frac{\tau\lambda}{2(\tau-\lambda)}
+\frac{\tau^{2}\lambda^{2}}{2(\tau-\lambda)^{2}}\qty[f(\tau^{-1})-f(\lambda^{-1})]
\notag \\ &\quad
+\frac{\tau^{2}\lambda}{(\tau-\lambda)^{2}}\qty[g(\tau^{-1})-g(\lambda^{-1})], \\
I_{2}(\tau, \lambda) &= -\frac{\tau\lambda}{2(\tau-\lambda)}\qty[f(\tau^{-1})-f(\lambda^{-1})].
\end{align}
The function $f(\tau)$ is found in Eq.~\eqref{eq: f_tau}, while $g(\tau)$ is defined as
\begin{align}
g(\tau) =
\begin{dcases}
\sqrt{\tau^{-1}-1}\arcsin(\sqrt{\tau}) &(\tau \geq 1), \\
\frac{\sqrt{1-\tau^{-1}}}{2}\qty[\ln{\frac{1+\sqrt{1-\tau^{-1}}}{1-\sqrt{1-\tau^{-1}}}}-i\pi] \quad &(\tau<1).
\end{dcases}
\label{eq: g_tau}
\end{align}
The ALP loop function is given by
\begin{align}
J_{\rm ALP}(\tau, \lambda)
&=
\frac{4\pi v}{\alpha_{\rm em}m_{h}^{2}}\qty(1-\frac{m_{Z}^{2}}{m_{h}^{2}})^{-1}
\Gamma_{hZ\gamma}^{1, {\rm 1PI}}(m_{Z}^{2},0,m_{h}^{2})_{\rm ALP}.
\label{eq: ALPloop_hZgam}
\end{align}
The term $\Delta_{\rm QCD}^{f}$ denotes QCD corrections to the quark-loop contributions.
In our analysis, it is included up to the next-to-leading order, and its expression is the same as that for the di-photon decay~\cite{Spira:1991tj}.

\subsection{$h \to aa$} \label{sec: haa_vetex}

The last channel in interest is $h \to aa$, \ie, a direct production of ALPs from the Higgs boson decay.
It proceeds via radiative corrections because the ALP is not coupled directly with the Higgs boson.
Therefore, the ALP contributions are obtained by the 1PI diagrams as
\begin{align}
\widehat{\Gamma}_{haa}(p_{1}^{2}, p_{2}^{2}, q^{2}) = \Gamma_{haa}^{\rm 1PI}(p_{1}^{2}, p_{2}^{2}, q^{2}).
\end{align}
Here, $p_{1}^{\mu}$ and $p_{2}^{\mu}$ are the incoming four-momenta of ALPs, and $q^{\mu}$ is the outgoing four-momentum of the Higgs boson (see the right panel of Fig.~\ref{fig: hToGamGam}).
The results are given in Appendix~\ref{app: three-point}.
As a result, the decay rate for $h \to aa$ is given by
\begin{align}
\Gamma(h\to aa) = \frac{1}{32\pi m_{h}}\sqrt{1-\frac{4m_{a}^{2}}{m_{h}^{2}}}
\abs{\Gamma_{haa}^{\rm 1PI}(m_{a}^{2}, m_{a}^{2}, m_{h}^{2})}^{2}.
\end{align}

\section{Results} \label{sec: numerical results}

In this section, we show numerical results of the ALP contributions to the Higgs boson decays.
For the channels whose final-states are composed only of the SM particles, the scaling factor of Higgs boson couplings is defined as
\begin{align}
\kappa_{XY} = \sqrt{\frac{\Gamma(h\to XY)}{\Gamma(h\to XY)_{\rm SM}}},
\end{align}
where the denominator is SM predictions, and the numerator includes both the SM and ALP contributions. 
In our analysis, the former is evaluated at the next-leading order by using the H-COUP program~\cite{Kanemura:2017gbi, Kanemura:2019slf}, and the ALP contributions are evaluated at the one-loop level.

Among the decay channels, the ALP corrections to $h \to f\bar f$ vanish at the one-loop level, as shown in Sec.~\ref{subsec: hToff}.
Hence, we focus on those into $\gamma, Z$, and $W$ as well as $h \to aa$ in the following.
As mentioned in Sec.~\ref{sec: ALP_model}, the ALP interactions are determined by two coupling constants. 
Thus, let us regard $g_{a\gamma\gamma}$ and $g_{aWW}$ as free parameters, while $g_{aZ\gamma}$ and $g_{aZZ}$ are represented by them. 
Also, the decay constant and cutoff scale are set as $f_{a}=1\TeV$ and $\Lambda=4\pi f_{a}$, respectively.

\begin{figure}[t]
 \centering
 \includegraphics[scale=0.45]{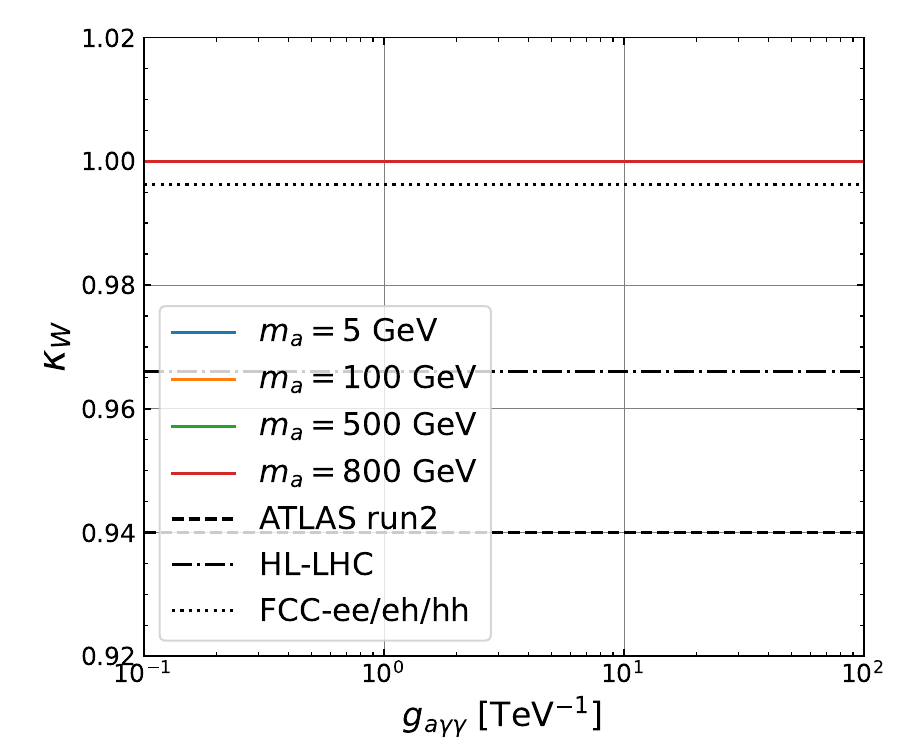} \hspace{7mm}
 \includegraphics[scale=0.45]{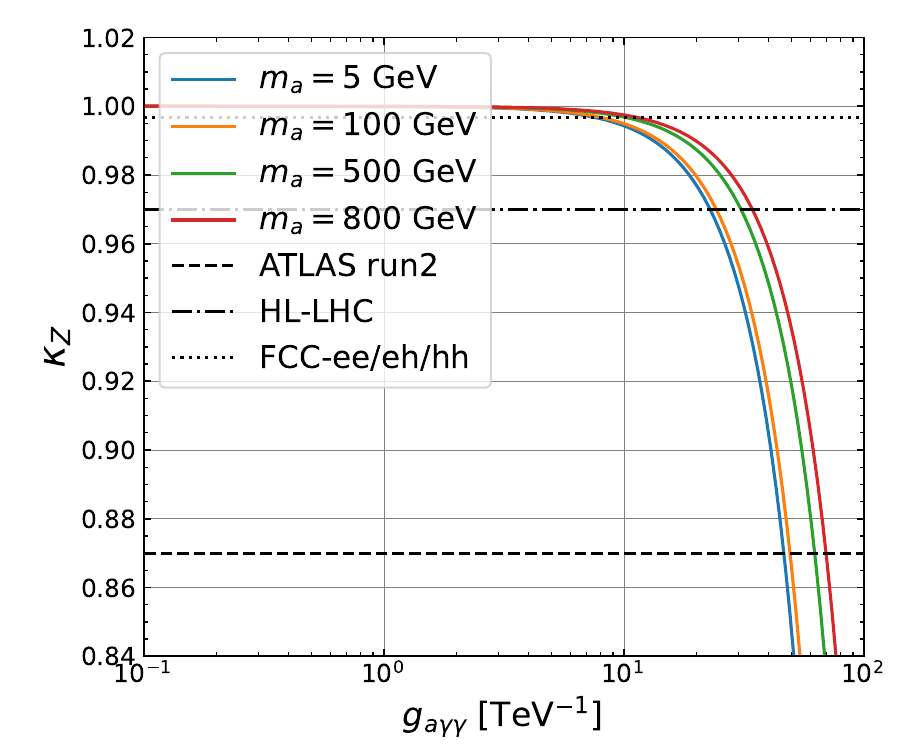}\\
 \includegraphics[scale=0.45]{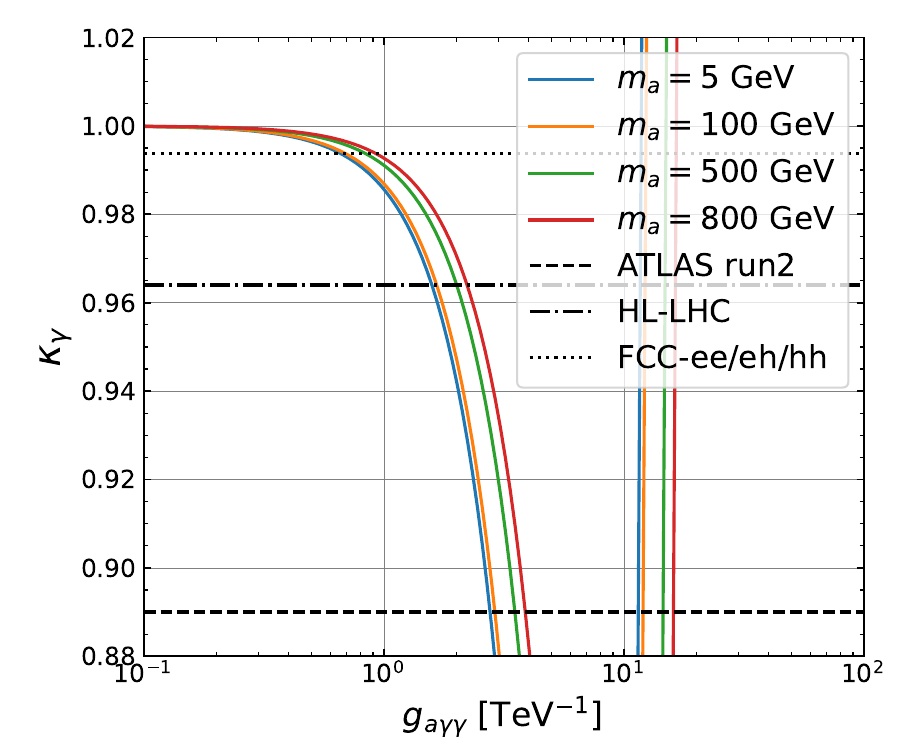} \hspace{7mm}
 \includegraphics[scale=0.45]{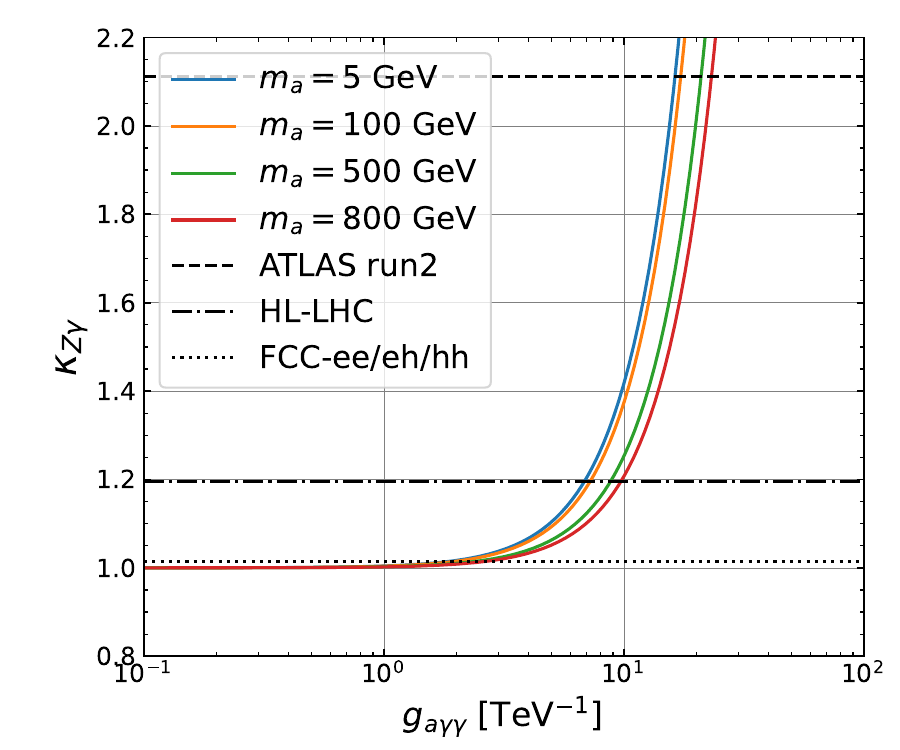}
 \caption{Scaling factors, $\kappa_{W}, \kappa_{Z}, \kappa_{\gamma}, \kappa_{Z\gamma}$, as a function of $g_{a\gamma\gamma}$ with fixing $g_{aW} = 0$.
 The ALP mass is chosen to be $m_a=5, 100, 500, 800\GeV$ (solid blue, yellow, green, and red lines, respectively).
 The horizontal dotted/dashed lines are experimental bounds and future sensitivities. }
 \label{fig: kappa_V_gaAA}
\end{figure}

In Fig.~\ref{fig: kappa_V_gaAA}, we show $\kappa_{W}, \kappa_{Z}, \kappa_{\gamma}, \kappa_{Z\gamma}$ as a function of $g_{a\gamma\gamma}$ with setting $g_{aWW} = 0$, where $\kappa_{X}$ denotes $\kappa_{XX}$.
The ALP mass is chosen as $m_a=5, 100, 500, 800\GeV$.
It is found that the ALP contributions are less sensitive to the value of $m_a$.
Moreover, there is no ALP contribution to $h \to WW^{*}$ because of $g_{aWW} = 0$.
On the other hand, the other channels receive corrections; they become effective for $g_{a\gamma\gamma} \gtrsim 1\TeV^{-1}$ in $h\to \gamma\gamma, Z\gamma$, and $10\TeV^{-1}$ in $h \to ZZ^*$.
It is noted that the ALP contribution interferes destructively with the SM one for $h\to \gamma\gamma, ZZ^*$, and its magnitude for $h\to \gamma\gamma$ becomes twice as large as or even larger than the SM contribution for $g_{a\gamma\gamma} \sim 10\TeV^{-1}$.

In the plots, experimental bounds and future prospects are shown by horizontal lines.
The dashed lines denote the bounds obtained by analyzing the full dataset of the ATLAS Run~II~\cite{ATLAS:2022vkf}.
The CMS experiment provides almost the same results~\cite{CMS:2022dwd}.
As for the prospects, the HL-LHC sensitivities~\cite{Cepeda:2019klc} are shown by the dot-dashed lines. 
We also show the dotted lines that represent far-future sensitivities expected by combining data from the FCC-ee/eh/hh experiments~\cite{deBlas:2019rxi}.

By comparing the ALP contributions with the experimental bound/sensitivities, it is found that $h \to \gamma\gamma$ provides the best probe to the ALP among the decay channels. 
The model is excluded for $g_{a\gamma\gamma} \gtrsim 2.8-3.9\TeV^{-1}$, depending on the ALP mass, except for the narrow region at $g_{a\gamma\gamma} \sim 10\TeV^{-1}$.
The HL-LHC experiment can probe the ALP with $g_{a\gamma\gamma} \gtrsim 1.6-2.2\TeV^{-1}$, and the sensitivity may reach $g_{a\gamma\gamma} = 0.7-0.9\TeV^{-1}$ at FCC-ee/eh/hh.

\begin{figure}[t]
 \centering
 \includegraphics[scale=0.45]{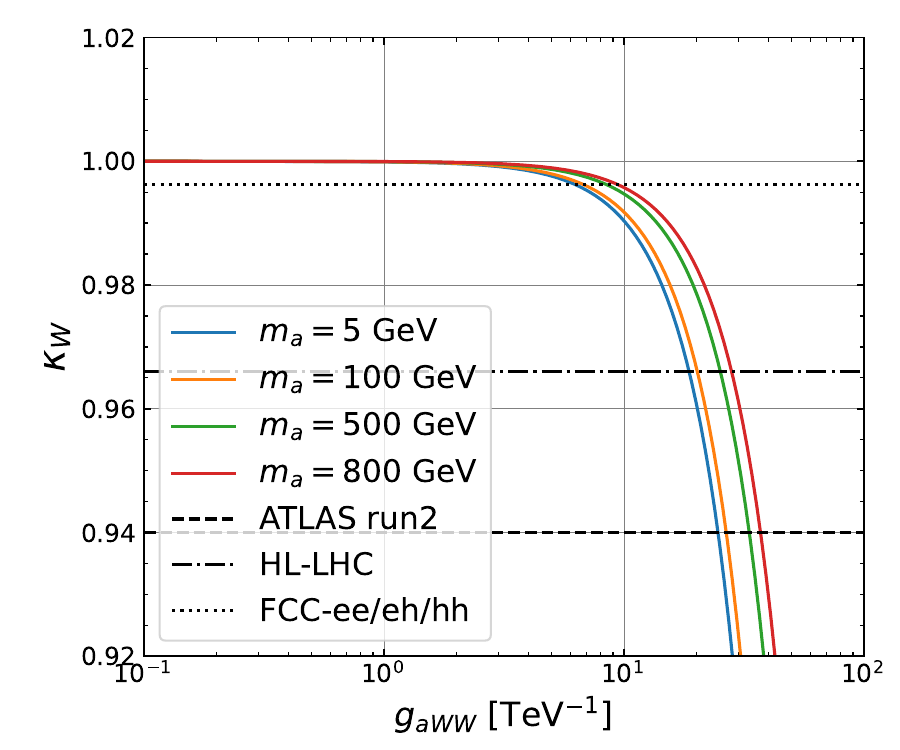} \hspace{7mm}
 \includegraphics[scale=0.45]{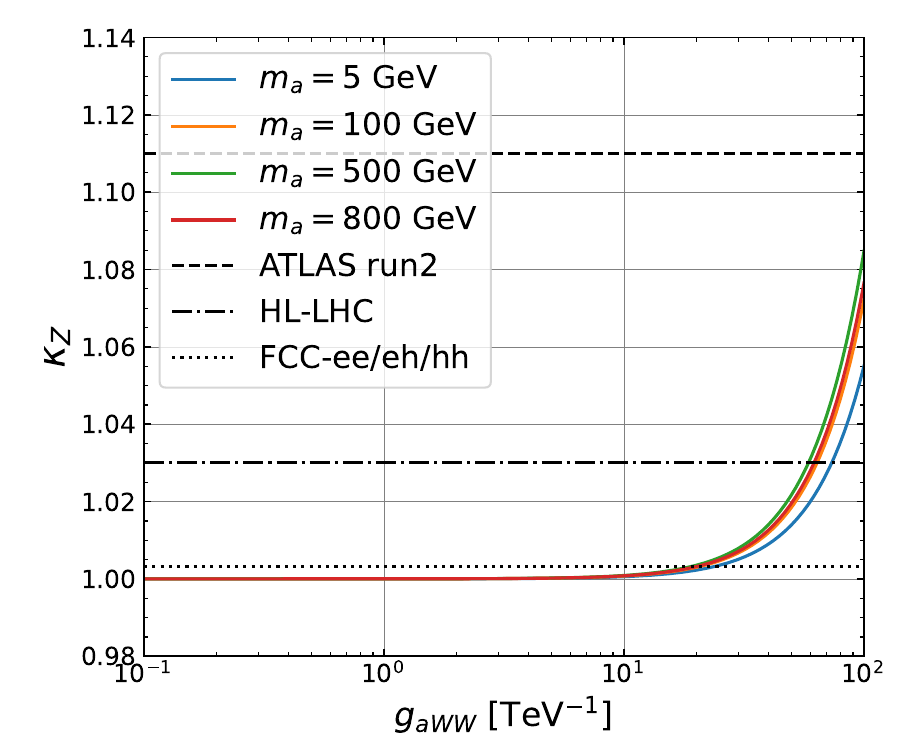}\\
 \includegraphics[scale=0.45]{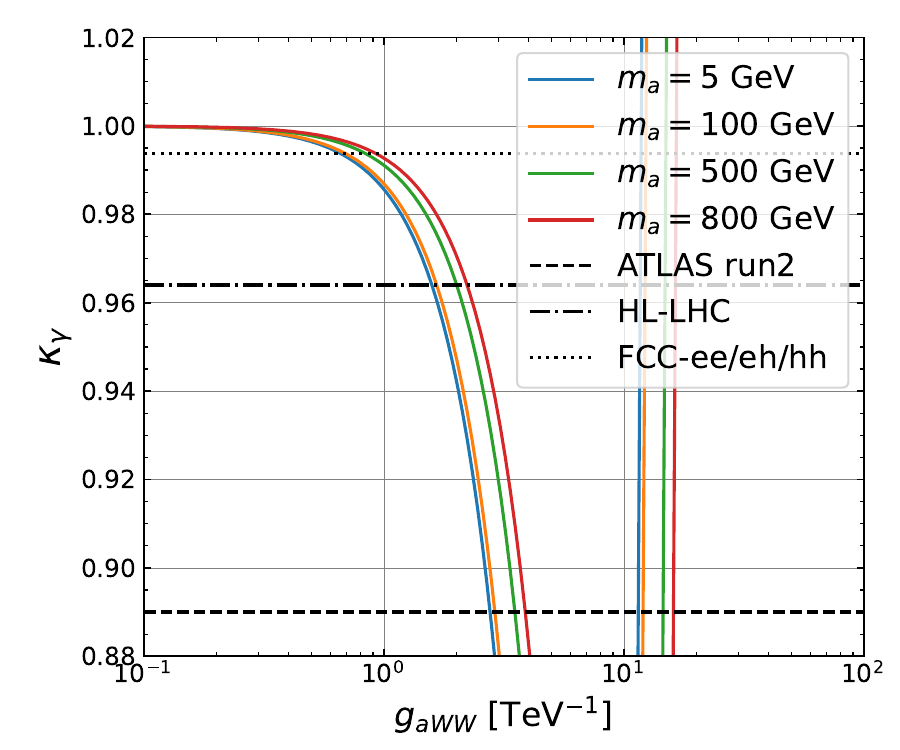} \hspace{7mm}
 \includegraphics[scale=0.45]{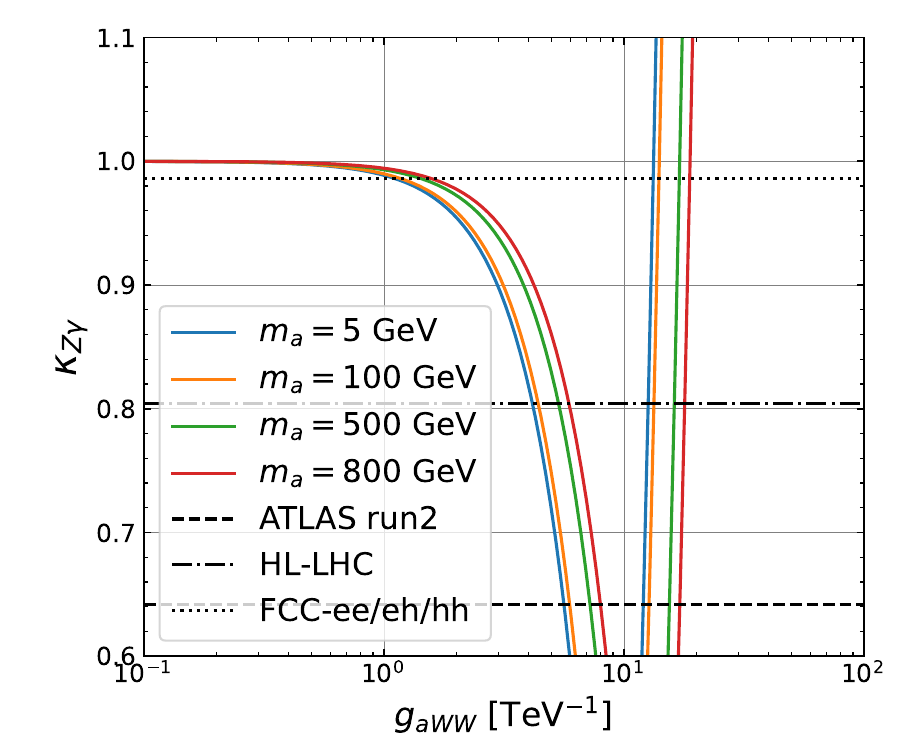}
 \caption{Same as Fig~\ref{fig: kappa_V_gaAA} but $g_{aWW}$ is varied with $g_{a\gamma\gamma} = 0$.}
 \label{fig: kappa_V_gaWW}
\end{figure}

In Fig.~\ref{fig: kappa_V_gaWW}, we study the case with $g_{aWW} \neq 0$ instead of $g_{a\gamma\gamma}$.
Here, $\kappa_{W}, \kappa_{Z}, \kappa_{\gamma}, \kappa_{Z\gamma}$ are shown as a function of $g_{aWW}$ with setting $g_{a\gamma\gamma} = 0$.
It is found that the ALP contributions are less sensitive to $m_a$ and become effective for $g_{aWW} \gtrsim 1\TeV^{-1}$ in $h\to \gamma\gamma, Z\gamma$ and $10\TeV^{-1}$ in $h\to ZZ^{*}, WW^{*}$.
The ALP contributions decrease the decay width of $h\to WW^{*}$ and increase that of $h\to ZZ^{*}$.
The difference between these two behaviors is understood by noticing that the $hZ\gamma$ vertex and the $Z\text{--}\gamma$ mixing give positive corrections in $h\to ZZ^{*}$.
On the other hand, the ALP contributions decrease the decay widths of $h\to \gamma\gamma$ and $h\to Z\gamma$ for $g_{aWW} \lesssim 10~{\rm TeV}^{-1}$.
However, the magnitude can become comparable to or larger than the SM corrections when $g_{aWW} \gtrsim 10~{\rm TeV}^{-1}$ and increase the decay widths compared with the SM predictions.

By comparing the ALP contributions with the experimental bound/sensitivities, it is found that $h \to \gamma\gamma$ provides the best probe to the ALP, similarly to the case with $g_{a\gamma\gamma} \neq 0$.
Since $g_{aZ\gamma} \propto (g_{aWW}-g_{a\gamma\gamma})$ is satisfied (see Eq.\eqref{eq: gaZgamma}), the values of $\kappa_{\gamma}$ in Fig.~\ref{fig: kappa_V_gaWW} become exactly the same as those in Fig.~\ref{fig: kappa_V_gaAA}.
Therefore, we obtain the same conclusion as the case with $g_{a\gamma\gamma} \neq 0$.

\begin{figure}[t]
 \centering
 \includegraphics[scale=0.36]{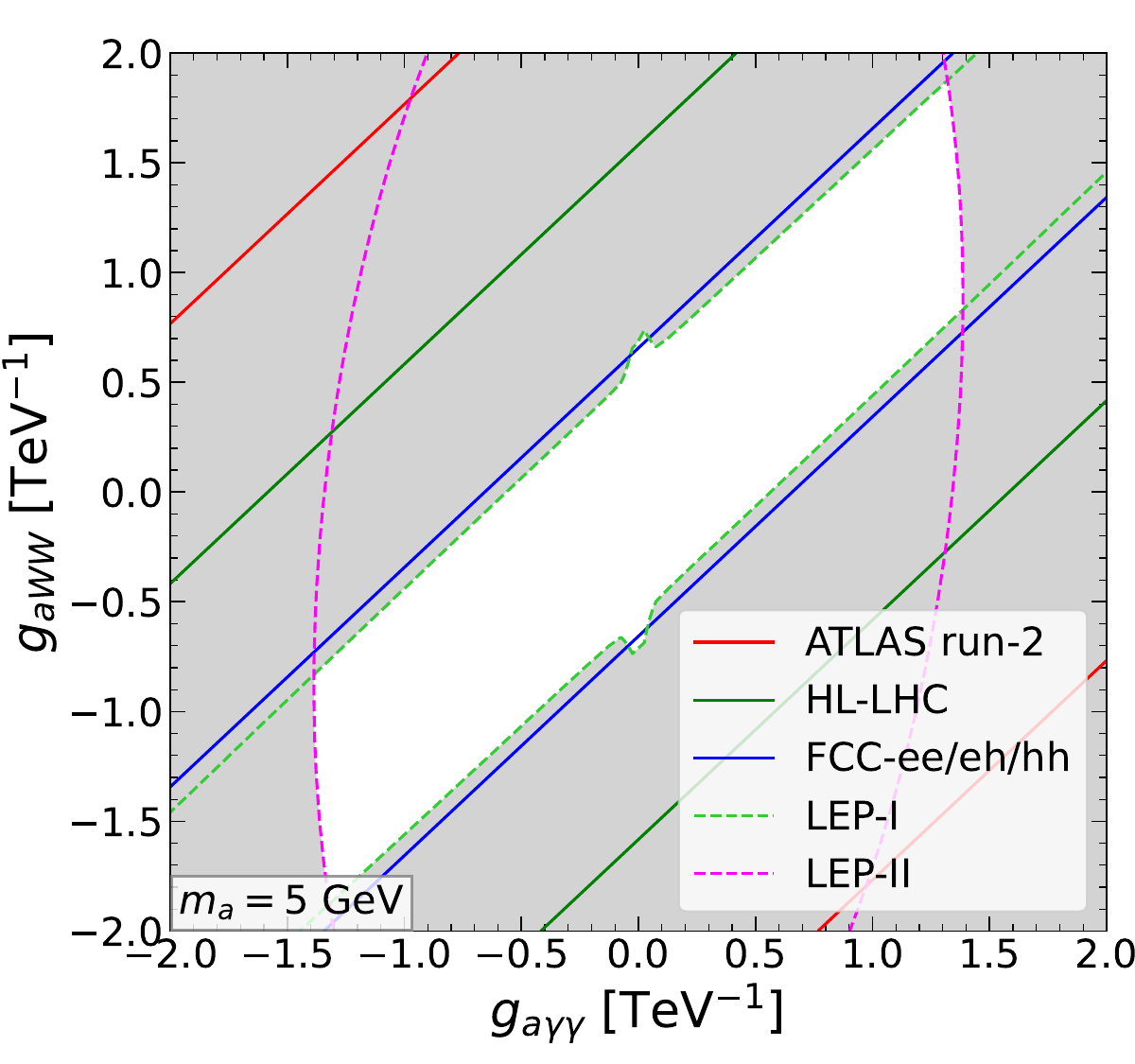}
 \includegraphics[scale=0.36]{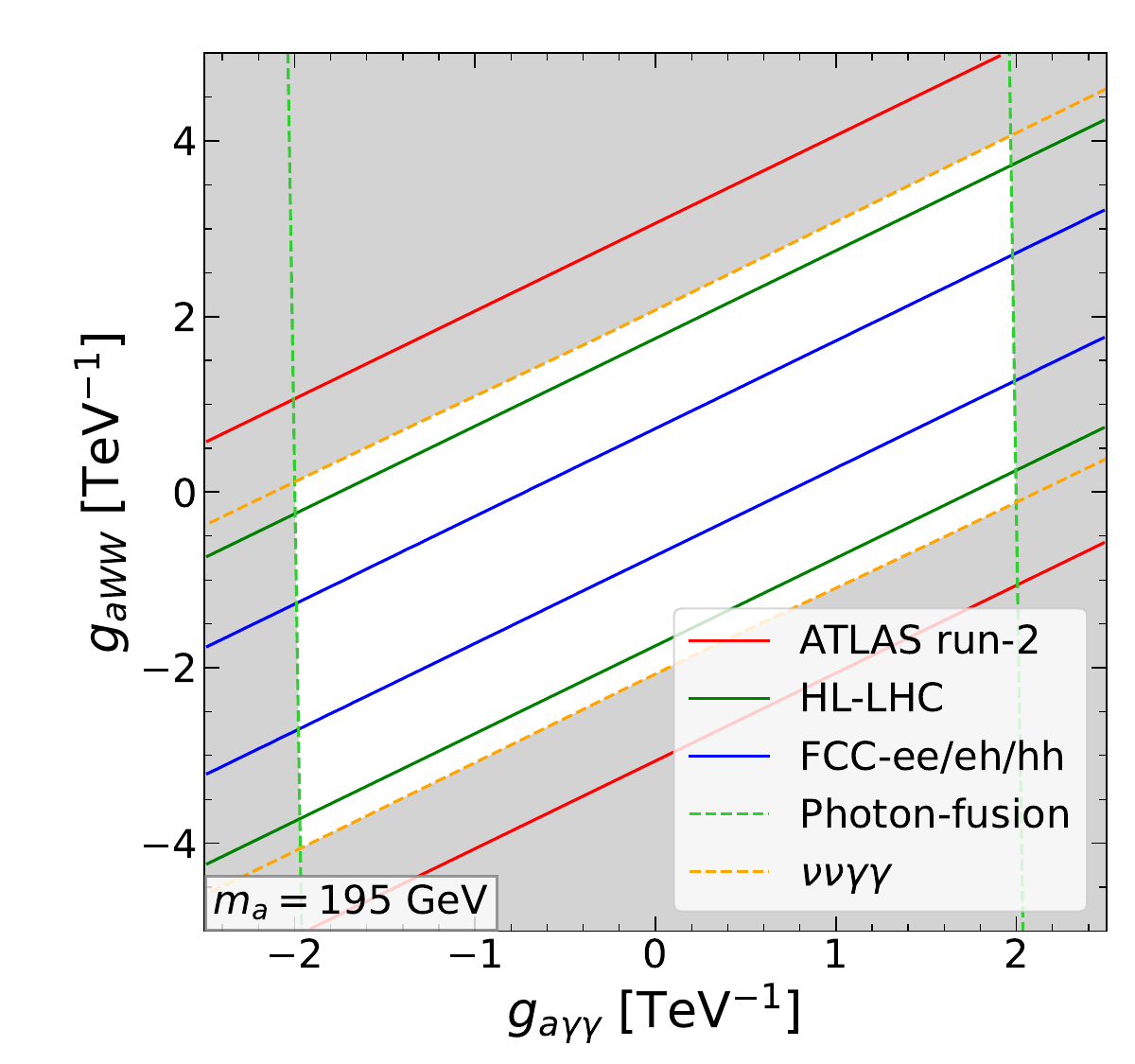}\\
 \includegraphics[scale=0.36]{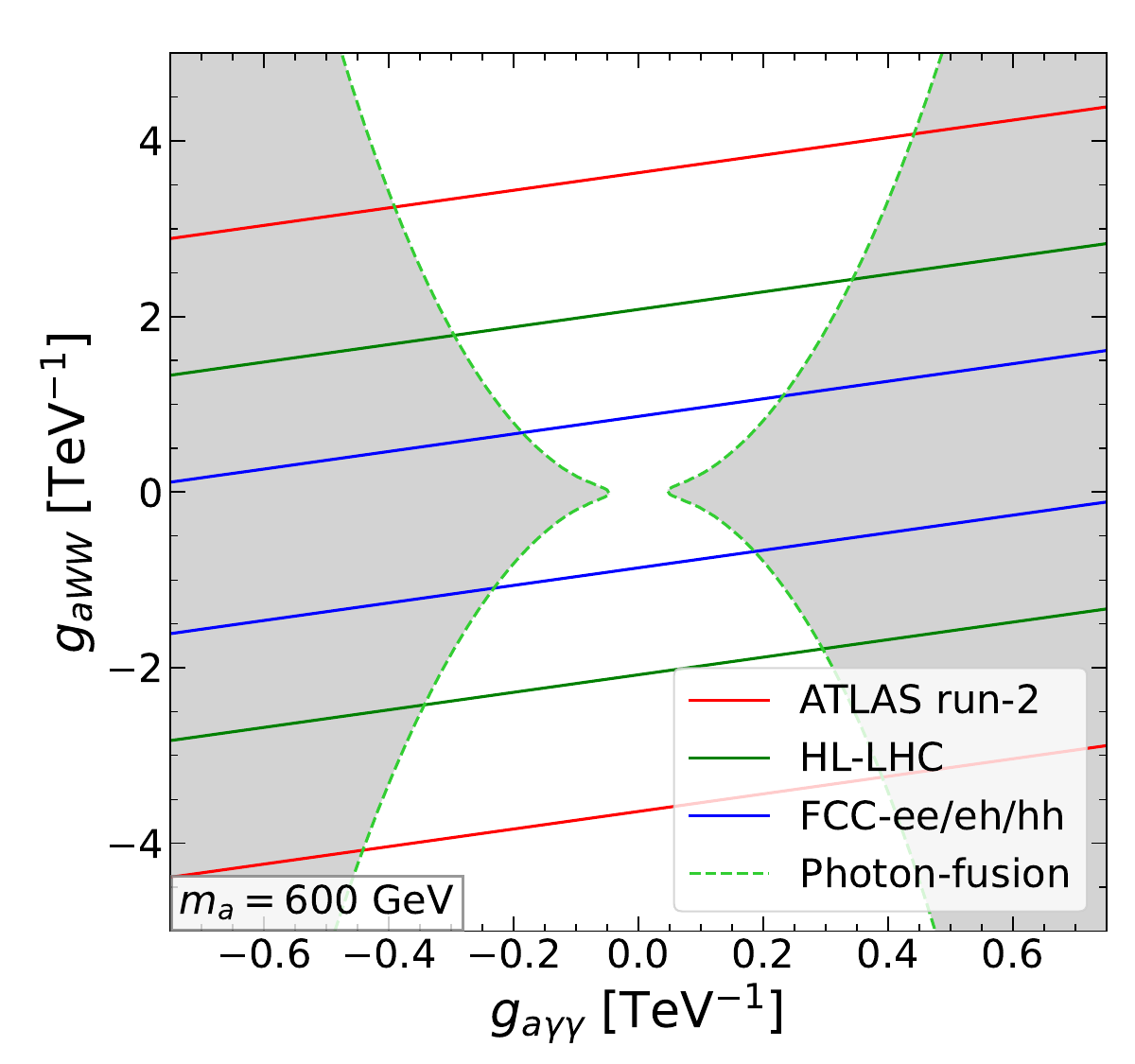} 
 \caption{Experimental bound (red) and future sensitivities (green and blue) of ALP contributions to $h \to \gamma\gamma$ (solid lines) on the $(g_{a\gamma\gamma}, g_{aWW})$ plane. The gray regions surrounded by dashed lines are excluded by the collider constraints.}
 \label{fig: kappa_A_2D}
\end{figure}

Let us next compare the Higgs results with collider constraints. 
In Fig.~\ref{fig: kappa_A_2D}, we show the collider limits on the $(g_{a\gamma\gamma}, g_{aWW})$ plane.
Here, we consider $m_a = 5, 195$, and $600\GeV$, where the constraints on $g_{a\gamma\gamma}$ is relatively weak (see the discussion in Ref.~\cite{Aiko:2023trb}).
The regions excluded by the collider limits are filled with gray color surrounded by dashed lines. 
For $m_a \lesssim 100\GeV$, the LEP experiments give strong bounds, while heavier ALPs are constrained by the LHC results. 
The details of the collider constraints are given in Ref.~\cite{Aiko:2023trb}.
It is commented that $h \to aa$ may have a sensitivity to the parameter space for $m_a = 5\GeV$.
However, we found that the effect is so weak that no lines appear in the parameter range of Fig.~\ref{fig: kappa_A_2D}.

In the plots, the experimental bound and future sensitivities to the ALP contributions to $h \to \gamma\gamma$ are shown by solid red (ATLAS Run~II bound), green (HL-LHC), and blue (FCC-ee/eh/hh) lines. 
For $m_a=5\GeV$, it is found that the $h \to \gamma\gamma$ sensitivities are almost weaker than the LEP constraints even at FCC-ee/eh/hh. 
By contrast, for $m_a=195\GeV$ the future sensitivity at HL-LHC can compete with the LHC constraint from the measurement of $pp \to \nu\nu\gamma\gamma$, and the model could be probed better by FCC-ee/eh/hh. 
For $m_a=600\GeV$, it is noticed that the collider constraints are relaxed especially for $g_{a\gamma\gamma} \sim 0$, and thus, $h \to \gamma\gamma$ provides a significant probe of the model. 

\begin{figure}[t]
 \centering
 \includegraphics[scale=0.6]{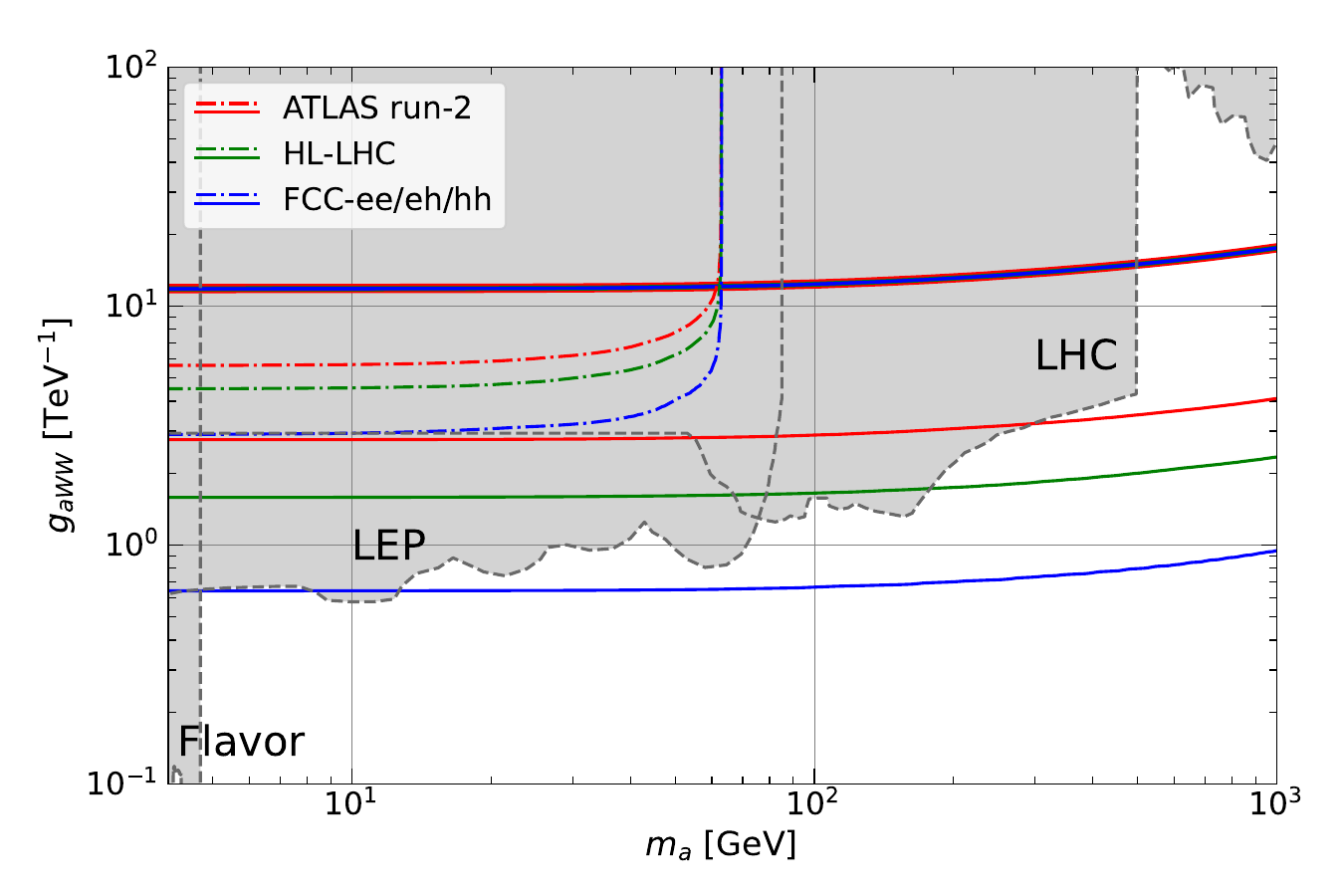} 
 \caption{Experimental bound (red) and future sensitivities (green and blue) of ALP contributions to $h \to \gamma\gamma$ (solid lines) and $h \to aa$ (dot-dashed lines) as a function of the ALP mass. 
 The gray regions surrounded by dashed lines are excluded by the collider constraints and/or flavor limits.}
 \label{fig: kappa_A_ma_gaWW}
\end{figure}

Since the model is much less constrained for $g_{a\gamma\gamma} \sim 0$, let us discuss the ALP mass dependence of the results for $g_{a\gamma\gamma} = 0$ and compare them with the collider constraints.
We also take flavor constraints into account for $m_a \lesssim 4.8\GeV$ (see Ref.~\cite{Aiko:2023trb} for details).
In Fig.~\ref{fig: kappa_A_ma_gaWW}, we show the experimental bound and future sensitivities on $g_{aWW}$ by the measurements of $\kappa_{\gamma}$ (solid lines) as a function of $m_a$.
It is found that the current bound on $\kappa_{\gamma}$ (red line) provides a stronger constraint than the colliders; the region of $g_{aWW} \gtrsim 3.2-4.1\TeV^{-1}$ is excluded for $m_a > 300\GeV$.
It is noted that there is a narrow allowed region above $g_{aWW} = 10\TeV^{-1}$ between the solid lines, where the ALP contribution is almost twice as large as the SM one with an opposite sign. 
As for future prospects, $g_{aWW} > 1.7-2.3\TeV^{-1}$ can be probed for $m_a > 180\GeV$ by HL-LHC (green line).
If FCC-ee/eh/hh would be constructed, the sensitivity could reach $g_{aWW} = 0.7-1\TeV^{-1}$ (blue line) and might compete with or even be better than the collider constraints at LEP and LHC.

In the figure, we also show the result of $h \to aa$, which is open kinematically for $m_a < m_h/2$.
The current bound and future sensitivities are shown by the dot-dashed red, green, and blue lines, respectively. 
It is found that the decay is less sensitive to the ALP model compared with $h \to \gamma\gamma$.

Before closing this section, let us comment on the case for $g_{aWW} = 0$. 
In Fig.~\ref{fig: kappa_A_ma_gaWW}, $g_{aWW}$ was varied with fixing $g_{a\gamma\gamma} = 0$. 
If $g_{a\gamma\gamma}$ is varied with $g_{aWW} = 0$, we obtain the same lines for $h \to \gamma\gamma$ as those in Fig.~\ref{fig: kappa_A_ma_gaWW} (see Eq.\eqref{eq: gaZgamma}).
However, the collider constraints are much more severe, and almost the whole parameter regions accessible by future $h \to \gamma\gamma$ measurements are already excluded. 
Similarly, if we consider the case for $g_{aZ\gamma} = 0$, \ie, $g_{a\gamma\gamma} = g_{aWW}$ is fixed, the ALP contributions to $h \to \gamma\gamma$ and $h \to Z\gamma$ are suppressed. 
Then, the ALP is probed by $h \to ZZ^*$ and $h \to WW^*$. 
Since the ALP contributions become effective only for $g_{a\gamma\gamma} \sim 10\TeV^{-1}$, such a parameter region is excluded by the collider constraints.

\section{Conclusion} \label{sec: conclusion}

In this paper, we have studied the Higgs boson decays in the ALP model.
The ALP is assumed to couple with the SM ${\rm SU(2)}_L$ and ${\rm U(1)}_Y$ gauge bosons.
We have provided the formulae for the ALP contributions to $h \to f\bar f$, $h \to \gamma\gamma$, $h \to Z\gamma$, $h \to ZZ$, and $h \to W^+W^-$ as well as $h \to aa$ at the one-loop level. 
It was found that those to $h \to f\bar f$ vanish at this level, while the other decay widths receive ALP contributions via radiative corrections.

Among the decay channels, it was concluded that $h \to \gamma\gamma$ can provide a significant probe of the model.
Compared with the collider constraints, the decay becomes relevant especially when the ALP is heavy and $g_{a\gamma\gamma}$ is suppressed.
For $g_{a\gamma\gamma} = 0$, it was shown that the current LHC result of $h \to \gamma\gamma$ excludes the ALP with $g_{aWW} \gtrsim 3.2-4.1\TeV^{-1}$ and $m_a > 300\GeV$.
In the future, the HL-LHC experiment can probe $g_{aWW} > 1.7-2.3\TeV^{-1}$ for $m_a > 180\GeV$.
Also, the sensitivity could reach $g_{aWW} = 0.7-1\TeV^{-1}$ at the FCC-ee/eh/hh experiments; it might compete with or even be better than the collider constraints on the direct ALP productions at LEP and LHC.
Moreover, $h \to \gamma\gamma$ has a better sensitivity than $h \to aa$.
Therefore, the precision measurements of $h \to \gamma\gamma$ could provide a powerful tool in searching for the ALP model.

\section*{Acknowledgements}
This work is supported by the Japan Society for the Promotion of Science (JSPS) Grant-in-Aid for Scientific Research on Innovative Areas (No.~22KJ3126 [MA]) and Scientific Research B (No.~21H01086 [ME]). 

\appendix
\section{ALP contributions} \label{app: 1PI}
In this section, we give the analytic expressions for the 1PI diagram contributions of the ALP to two-point and three-point functions.
The calculations were performed in the $R_{\xi}$ gauge by using \texttt{FeynCalc}~\cite{Mertig:1990an, Shtabovenko:2016sxi, Shtabovenko:2020gxv} and \texttt{FeynArts}~\cite{Hahn:2000kx} with a model file generated with \texttt{FeynRules}~\cite{Alloul:2013bka}.
We have checked that all the results are shown to be gauge invariant.

\subsection{Loop functions} \label{app: loop_function}
The loop integrals with a tensor structure can be decomposed into linear combinations of the metric tensor and a set of the momenta with scalar functions~\cite{Passarino:1978jh}.
The loop integrals are defined as
\begin{align}
A_{0}
&=
\int\frac{\overline{\dd^{d}k}}{i\pi^{2}}
\frac{1}{N_{0}}, \\
B_{0}
&=
\int\frac{\overline{\dd^{d}k}}{i\pi^{2}}
\frac{1}{N_{0}N_{1}}, \\
[C_{0}, C^{\mu}, C^{\mu\nu}]
&=
\int\frac{\overline{\dd^{d}k}}{i\pi^{2}}
\frac{[1,k^{\mu},k^{\mu}k^{\nu}]}{N_{0}N_{1}N_{2}},
\end{align}
where $d=4-2\epsilon$ and $\overline{\dd^{d}k}=\Gamma(1-\epsilon)(\pi\mu^{2})^{\epsilon}\dd^{d}k$ in the
$\overline{\rm MS}$ regularization~\cite{Hagiwara:1994pw}.
The propagator factor is defined as $N_{i} = (k+r_{i})^{2}-m_{i}^{2}+i\epsilon$ where $r_{0}=0$ and $r_{N}=\sum_{i=1}^{N}p_{i}$ with incoming four-momenta $p_{i}$.
We take $\mu=\Lambda$ in this paper.
The $C^{\mu}$ and $C^{\mu\nu}$ functions are decomposed as~\cite{Denner:1991kt}
\begin{align}
C^{\mu} &= \sum_{i=1}^{2}r_{i}^{\mu}C_{i}(p_{1}^{2}, p_{2}^{2}, q^{2}; m_{0}^{2}, m_{1}^{2}, m_{2}^{2}), \\
C^{\mu\nu} &= \qty[g^{\mu\nu} C_{00}+\sum_{i,j=1}^{2}r_{i}^{\mu}r_{j}^{\nu}C_{ij}](p_{1}^{2}, p_{2}^{2}, q^{2}; m_{0}^{2}, m_{1}^{2}, m_{2}^{2}).
\end{align}
In this paper, we use \texttt{LoopTools}~\cite{Hahn:1998yk} for the evaluation of scalar functions.
Also, we use the following abbreviations for scalar functions.
\begin{align}
A_{0}(X) &= A_{0}(m_{X}^{2}), \\
B_{0}(p^{2}; X, Y) &= B_{0}(p^{2}; m_{X}^{2}, m_{Y}^{2}), \\
C_{i,\, ij}(X, Y, Z) &= C_{i,\, ij}(p_{1}^{2}, p_{2}^{2}, q^{2}; m_{X}^{2}, m_{Y}^{2}, m_{Z}^{2}).
\end{align}

\subsection{Two-point functions} \label{app: two-point}

The ALP contributions to two-point functions of the EW gauge boson are given as~\cite{Aiko:2023trb}
\begin{align}
16\pi^{2}\Pi_{WW}^{{\rm 1PI}, T}(p^{2})_{\rm ALP}
&=
\frac{1}{18}g_{aWW}^{2}F(p^{2}; a, W), \\
16\pi^{2}\Pi_{\gamma\gamma}^{{\rm 1PI}, T}(p^{2})_{\rm ALP}
&=
\frac{1}{18}\qty[g_{a\gamma\gamma}^{2} F(p^{2}; a, \gamma) + g_{aZ\gamma}^2 F(p^{2}; a, Z)], \\
16\pi^{2}\Pi_{Z\gamma}^{{\rm 1PI}, T}(p^{2})_{\rm ALP}
&=
\frac{1}{18}\qty[g_{a\gamma\gamma}g_{aZ\gamma} F(p^{2}; a, \gamma) + g_{aZ\gamma}g_{aZZ} F(p^{2}; a, Z)], \\
16\pi^{2}\Pi_{ZZ}^{{\rm 1PI}, T}(p^{2})_{\rm ALP}
&=
\frac{1}{18}\qty[g_{aZ\gamma}^{2} F(p^{2}; a, \gamma) + g_{aZZ}^2 F(p^{2}; a, Z)],
\end{align}
where the loop function is defined as
\begin{align}
F(p^{2}; a, V) &= 
3\qty[p^{2}-\qty(m_{a}+m_{V})^2]\qty[p^{2}-\qty(m_{a}-m_{V})^2]
\qty[B_0(p^{2}; a, V)-B_{0}(0; a, V)]
\notag \\ &\quad
-3p^{2}\qty[A_{0}(a)+A_{0}(V)+\qty(2m_{a}^{2}+2m_{V}^{2}-p^{2})B_{0}(0; a, V)]
\notag \\ &\quad
+7p^{2}(3m_{a}^{2}+3m_{V}^{2}-p^{2}).
\label{eq: F_function}
\end{align}
The derivative of the loop function is explicitly shown as
\begin{align}
F'(p^{2}; a, V) &= 
6(p^{2}-m_{a}^{2}-m_{V}^{2})
\qty[B_{0}(p^{2}; a, V)-B_{0}(0; a, V)]
\notag \\ &\quad
+3\qty[p^{2}-\qty(m_{a}+m_{V})^2]\qty[p^{2}-\qty(m_{a}-m_{V})^2]
B'_{0}(p^{2}; a, V)
\notag \\ &\quad
-3\qty[A_{0}(a)+A_{0}(V)+2\qty(m_{a}^{2}+m_{V}^{2}-p^{2})B_{0}(0; a, V)]
\notag \\ &\quad
+7(3m_{a}^{2}+3m_{V}^{2}-2p^{2}).
\label{eq: Fprime_function}
\end{align}

\subsection{Three-point functions} \label{app: three-point}
The ALP contributions to the form factors of the $hWW$ vertices are given by
\begin{align}
16\pi^{2}\Gamma^{1, {\rm 1PI}}_{hWW}(p_{1}^{2}, p_{2}^{2}, q^{2})_{\rm ALP}
&=
\frac{gm_{W}g_{aWW}^{2}}{8}C_{hVV1}^{VSV}(W, a, W)_{\rm ALP}, \\
16\pi^{2}\Gamma^{2, {\rm 1PI}}_{hWW}(p_{1}^{2}, p_{2}^{2}, q^{2})_{\rm ALP}
&=
\frac{gm_{W}^{3}g_{aWW}^{2}}{2}C_{hVV2}^{VSV}(W, a, W)_{\rm ALP}, \\
\Gamma^{3, {\rm 1PI}}_{hWW}(p_{1}^{2}, p_{2}^{2}, q^{2})_{\rm ALP}
&=
0.
\end{align}
where the loop functions are obtained as
\begin{align}
&C_{hVV1}^{VSV}(X, a, X)_{\rm ALP}
\notag \\ &
=
(p_{1}^{2}+p_{2}^{2}-3q^{2}-2m_{a}^{2}+2m_{X}^{2})B_{0}(q^{2}; X, X)
\notag \\ &\quad
+2(p_{1}^{2}+m_{a}^{2}-m_{X}^{2})B_{0}(p_{1}^{2}; a, X)
+2(p_{2}^{2}+m_{a}^{2}-m_{X}^{2})B_{0}(p_{2}^{2}; a, X)
\notag \\ &\quad
-2\qty[p_{1}^{2}p_{2}^{2}-p_{1}^{2}(m_{a}^{2}+m_{X}^{2})-p_{2}^{2}(m_{a}^{2}+m_{X}^{2})+2m_{a}^{2}q^{2}+\qty(m_{a}^{2}-m_{X}^{2})^{2}]C_{0}(X, a, X)
\notag \\ &\quad
+4(q^{2}-p_{1}^{2}-p_{2}^{2})C_{00}(X, a, X)
+4(q^{2}-p_{1}^{2}-p_{2}^{2})-2A_{0}(a)+2A_{0}(X), \\
&C_{hVV2}^{VSV}(X, a, X)_{\rm ALP}
\notag \\ &
=
B_{0}(q^{2}; X, X)
+2m_{a}^{2}C_{0}(X, a, X)
\notag \\ &\quad
+(p_{1}^{2}+m_{a}^{2}-m_{X}^{2})C_{1}(a, X, X)
+(p_{2}^{2}+m_{a}^{2}-m_{X}^{2})C_{1}(a, X, X)
\notag \\ &\quad
+\qty(p_{1}^{2}+p_{2}^{2}-q^{2})C_{12}(a, X, X)
-2.
\end{align}
The ALP contributions to the form factors of the $hZZ$ vertices are given by
\begin{align}
16\pi^{2}\Gamma^{1, {\rm 1PI}}_{hZZ}(p_{1}^{2}, p_{2}^{2}, q^{2})_{\rm ALP}
&=
\frac{g_{Z}m_{Z}g_{aZZ}^{2}}{8}C_{hVV1}^{VSV}(Z, a, Z)_{\rm ALP}, \\
16\pi^{2}\Gamma^{2, {\rm 1PI}}_{hZZ}(p_{1}^{2}, p_{2}^{2}, q^{2})_{\rm ALP}
&=
\frac{g_{Z}m_{Z}^{3}g_{aZZ}^{2}}{2}C_{hVV2}^{VSV}(Z, a, Z)_{\rm ALP}, \\
\Gamma^{3, {\rm 1PI}}_{hZZ}(p_{1}^{2}, p_{2}^{2}, q^{2})_{\rm ALP}
&=
0.
\end{align}
The ALP contributions to the form factors of the $hZ\gamma$ vertices are given by
\begin{align}
16\pi^{2}\Gamma^{1, {\rm 1PI}}_{hZ\gamma}(p_{1}^{2}, p_{2}^{2}, q^{2})_{\rm ALP}
&=
\frac{g_{Z}m_{Z}g_{aZ\gamma}g_{aZZ}}{8}C_{hVV1}^{VSV}(Z, a, Z)_{\rm ALP}, \\
16\pi^{2}\Gamma^{2, {\rm 1PI}}_{hZ\gamma}(p_{1}^{2}, p_{2}^{2}, q^{2})_{\rm ALP}
&=
\frac{g_{Z}m_{Z}g_{aZ\gamma}g_{aZZ}}{2}C_{hVV2}^{VSV}(Z, a, Z)_{\rm ALP}, \\
\Gamma^{3, {\rm 1PI}}_{hZ\gamma}(p_{1}^{2}, p_{2}^{2}, q^{2})_{\rm ALP}
&=
0.
\end{align}
The ALP contributions to the form factors of the $h\gamma\gamma$ vertices are given by
\begin{align}
16\pi^{2}\Gamma^{1, {\rm 1PI}}_{h\gamma\gamma}(p_{1}^{2}, p_{2}^{2}, q^{2})_{\rm ALP}
&=
\frac{g_{Z}m_{Z}g_{aZ\gamma}^{2}}{8}C_{hVV1}^{VSV}(Z, a, Z)_{\rm ALP}, \\
16\pi^{2}\Gamma^{2, {\rm 1PI}}_{h\gamma\gamma}(p_{1}^{2}, p_{2}^{2}, q^{2})_{\rm ALP}
&=
\frac{g_{Z}m_{Z}g_{aZ\gamma}^{2}}{2}C_{hVV2}^{VSV}(Z, a, Z)_{\rm ALP}, \\
\Gamma^{3, {\rm 1PI}}_{h\gamma\gamma}(p_{1}^{2}, p_{2}^{2}, q^{2})_{\rm ALP}
&=
0.
\end{align}
The gauge boson contributions to the $haa$ vertex are given by
\begin{align}
16\pi^{2}\Gamma^{\rm 1PI}_{haa}(p_{1}^{2}, p_{2}^{2}, q^{2})_{\rm ALP}
&=
\frac{g m_{W} g_{aWW}^{2}}{2}C_{haa}^{VVV}(W, W)
+\frac{g_{Z}m_{Z} g_{aZZ}^{2}}{4}C_{haa}^{VVV}(Z, Z)
\notag \\ &\quad
+\frac{g_{Z}m_{Z} g_{aZ\gamma}^{2}}{4}C_{haa}^{VVV}(Z, \gamma),
\end{align}
where the loop function is obtained as
\begin{align}
&C_{haa}^{VVV}(V_{1}, V_{2})
\notag \\ &
=
2\qty[p_{1}^{2}p_{2}^{2}+2m_{V_{2}}^{2}q^{2}-(m_{V_{1}}^{2}+m_{V_{2}}^{2})(p_{1}^{2}+p_{2}^{2})+(m_{V_{1}}^{2}-m_{V_{2}}^{2})^{2}]
C_{0}(V_{1}, V_{2}, V_{1})
\notag \\ &\quad
+(3q^{2}-p_{1}^{2}-p_{2}^{2}-2m_{V_{1}}^{2}+2m_{V_{2}}^{2})B_{0}(q^{2}; V_{1}, V_{1})
\notag \\ &\quad
-2(p_{1}^{2}-m_{V_{1}}^{2}+m_{V_{2}}^{2})B_{0}(p_{1}^{2}; V_{1}, V_{2})
-2(p_{2}^{2}-m_{V_{1}}^{2}+m_{V_{2}}^{2})B_{0}(p_{2}^{2}; V_{1}, V_{2})
\notag \\ &\quad
-2A_{0}(V_{1})+2A_{0}(V_{2})
-9(q^{2}-p_{1}^{2}-p_{2}^{2}).
\end{align}

\bibliographystyle{utphys28mod}
\bibliography{references}
\end{document}